\documentclass[aps,superscriptaddress]{revtex4}
\usepackage{amssymb,amsmath,epsfig}
\usepackage{subfigure}
\usepackage[colorlinks=true, pdfstartview=FitV, linkcolor=blue, citecolor=red, urlcolor=magenta, breaklinks=true]{hyperref}
\begin{document}
\title{Absorption, scattering and shadow by a noncommutative black hole with global monopole}

\author{M. A. Anacleto}\email{anacleto@df.ufcg.edu.br}
\affiliation{Departamento de F\'{\i}sica, Universidade Federal de Campina Grande
Caixa Postal 10071, 58429-900 Campina Grande, Para\'{\i}ba, Brazil}

\author{F. A. Brito\footnote{\!\!Corresponding author}}\email{fabrito@df.ufcg.edu.br}
\affiliation{Departamento de F\'{\i}sica, Universidade Federal de Campina Grande
Caixa Postal 10071, 58429-900 Campina Grande, Para\'{\i}ba, Brazil}
\affiliation{Departamento de F\'isica, Universidade Federal da Para\'iba, 
Caixa Postal 5008, 58051-970 Jo\~ao Pessoa, Para\'iba, Brazil}
 
\author{J. A. V. Campos}\email{joseandrecampos@gmail.com}
\affiliation{Departamento de F\'isica, Universidade Federal da Para\'iba, 
Caixa Postal 5008, 58051-970 Jo\~ao Pessoa, Para\'iba, Brazil}

\author{E. Passos}\email{passos@df.ufcg.edu.br}
\affiliation{Departamento de F\'{\i}sica, Universidade Federal de Campina Grande
Caixa Postal 10071, 58429-900 Campina Grande, Para\'{\i}ba, Brazil}

\begin{abstract} 
In this paper, we investigate the process of massless scalar wave scattering due to a noncommutative black hole with a global monopole through the partial wave method. We computed the cross section of differential scattering and absorption at the low frequency limit. 
We also calculated, at the high frequency limit, the absorption and the shadow radius by the null geodesic method.
We showed that noncommutativity causes a reduction in the differential scattering/absorption cross
section and shadow radius, while the presence of the global monopole has the effect of increasing the value of such quantities.
In the limit of the null mass parameter, we verify that the cross section of differential scattering, absorption and shadow ray approach to a non-zero value proportional to a minimum mass.

\end{abstract}

\maketitle
\pretolerance10000

\section{Introduction}

Hawking~\cite{Hawking:1971vc} black hole radiation emission studies revealed a physically relevant possibility for us to enter the universe of quantum gravity. A very important point raised by Hawking is related to the final stage of black hole
evaporation, so that when the mass of the black hole approaches the Planck mass, a quantum theory of gravity is needed. 
However, stimulated by the need to study the structure of spacetime on the Planck scale, many other approaches have
been proposed considering noncommutative geometry~\cite{Padmanabhan:2003gd,Szabo:2006wx,Ansoldi:2006vg}. 
As with position and moment uncertainty in conventional quantum mechanics, we can find a noncommutative relationship in Einstein theory of general relativity. 
Thus, position measurements can fail to commute by suggesting a noncommutative manifold~\cite{Nicolini:2008aj}. 
Studies as proposed by Nicolini~\cite{Nicolini:2005vd} showed that noncommutativity can be implemented in general relativity by modifying the matter source. This can be done by modifying the mass density by replacing a Dirac delta function with a Gaussian distribution of minimum width $ \sqrt{\theta} $ given by $ \rho_{\theta}(r)=M(4\pi\theta)^{-3/2}\exp{(-r^2/4\theta)} $, 
or even considering replacing by a Lorentzian
distribution~\cite{Nozari:2008rc} tanking the following form
$\rho_{\theta}(r)=M\sqrt{\theta}\pi^{-3/2}(r^2+\pi\theta)^{-2}$,
where $ \theta $ is the noncommutative parameter and $M$ is the total mass that due to the uncertainty of noncommutativity is diffused throughout the linear region of size $ \sqrt{\theta} $. 
Besides the noncommutativity, the study of topological defects has been used in different areas of
physics as a way to improve understanding the behavior of the early Universe. 

It is well known in the literature that global monopoles are types of topological defects that are formed by a spontaneous symmetry breaking of global symmetry and that a large part of its energy is concentrated into a point more specifically in the center~\cite{Vilenkin:2000jqa} and, a charge of this nature is characterized by a spontaneous symmetry breaking of original global $O(3)$ symmetry to $U(1)$~\cite{Kibble:1976sj}. 
In recent years many authors have devoted themselves to studying black holes 
with global monopoles~\cite{BezerradeMello:1996si,Pitelli:2009kd,Sharif:2015kna,Shaikh:2019fpu,Haroon:2019new}, initiated by Barriola and Vilenkin~\cite{Barriola:1989hx} which was the first study considering a metric of black hole with global static monopoles.
In~\cite{Anacleto:2017kmg}, the differential scattering cross section and the absorption of a black hole with global monopole in $ f(R) $ gravity  have been computed by the partial wave method.

In the present work, we have the main purpose of exploring the effect of quantum gravity corrections that contribute
to the process of massless scalar wave scattering by a noncommutative black hole with a global monopole. 
The problem of scattering a massless scalar wave across a black hole has been constantly studied
by several authors~\cite{Townsend:1997ku,Futterman:1988ni,Li:2022wzi,Xing:2022emg,Bisnovatyi-Kogan:2022ujt,Zeng:2021dlj,Mourad:2021qgo,Jha:2021bue,Li:2022jda,Gogoi:2022wyv,Karmakar:2022idu,Lobos:2022jsz,Tsupko:2022yzg,Zeng:2022fdm,Heydari-Fard:2021qdc,Heydari-Fard:2021pjc,Khodadi:2021gbc,Fathi:2020sfw,Chen:2022ngd}, and many works considering the low frequency limit 
($GM \ll 1$) can be found in the literature~\cite{Matzner:1977dn,Westervelt:1971pm,Peters:1976jx,Sanchez:1976fcl,Sanchez:1976xm,Sanchez:1977si,Sanchez:1977vz,DeLogi:1977dp,Doran:2001ag,Dolan:2007ut,Crispino:2009ki,Starobinskil:1974nkd}. The use of the partial wave method can also be found in several different contexts~\cite{Gibbons:1975kk,Page:1976df,Churilov1973,Moura:2011rr,Jung:2004yh,Jung:2004yn,Doran:2005vm,Dolan:2006vj,Castineiras:2007ma,Benone:2014qaa,Marinho:2016ixt,Das:1996we,Macedo:2016yyo,deOliveira:2018kcq,Hai:2013ara}. 
Besides, this method has also been applied to examine scalar wave scattering by acoustic black
holes~\cite{Crispino:2007zz,Dolan:2009zza,Oliveira:2010zzb,Dolan:2011zza,Anacleto:2012ba,Anacleto:2012du,Dolan:2012yc,Anacleto:2015mta,Anacleto:2016ukc} 
and also to exploring the scattering of scalar waves by a noncommutative BTZ black hole~\cite{Anacleto:2014cga}.

In previous studies~\cite{Anacleto:2019tdj}, we investigated the effect of noncommutativity in
black hole by considering a Lorentzian smeared mass distribution on the scattering problem, where we find a differential
scattering cross section for small angles given approximately by 
$ d\sigma/d\Omega\approx 16G^2M^2/\vartheta^2[1+(4/M)\sqrt{\theta/\pi}] $.
In the present work, we will see that the presence of noncommutativity and the global monopole changes the behavior of the differential scattering cross section in a nontrivial way.
In addition, we also saw in~\cite{Anacleto:2019tdj} that for the low frequency limit the absorption cross section is always proportional to the
area of the event horizon of the noncommutative Schwarzschild black hole 
$ \sigma\approx 16G^2M^2/\vartheta^2[1+(2/M)\sqrt{\theta/\pi}]^2 $.
We will see that the absorption cross section increases when we add the influence of the global monopoles on the
geometry. 
At the small mass limit $ M \to0 $, we have a result that differs from studies done for other geometries~\cite{Anacleto:2018acl,Anacleto:2020lel,Anacleto:2020zhp}, and at this limit the absorption cross section has a direct relationship with the mass $ M $. 
Therefore, starting from a Lorentzian distribution, analytical results are more easily examined than using a Gaussian distribution where the analysis is done numerically. 
In~\cite{Anacleto:2020efy,Anacleto:2020zfh}, employing the Lorentzian distribution, the authors have found logarithmic corrections for entropy as well as the condition for the black hole remnant.
In~\cite{Campos:2021sff}, we have investigated the quasinormal modes and the shadow radius of a noncommutative Schwarzschild black hole. 
We show that, in the zero-mass limit, the shadow ray does not vanish, being proportional to a minimum mass for the finite noncommutative parameter and the black hole becomes a black hole remnant.
Therefore, in the present work, we aim to explore the effect of quantum gravity corrections that contribute to the process of massless scalar wave scattering by a noncommutative black hole with a global monopole. 
We will also apply the null geodesic method to calculate the shadow radius in order to verify the influence of noncommutativity and the global monopole.

The paper is organized as follows. In Sec.~\ref{sc2} we consider the influence of noncommutativity and the global monopole on the Schwarzschild black hole metric to determine the differential cross section and absorption. 
In Sec.~\ref{ng} we apply the null geodesic method to compute the shadow of the noncommutative black hole with a global monopole.
In Sec.~\ref{conc} we make our final considerations.
Here we adopt the natural units $ \hbar=c=k_B=G=1$.

\section{Noncommutative Black Hole with a global monopole}
\label{sc2}
In this section, we introduce the noncommutative black hole with a global monopole for the purpose of determining the differential scattering cross section and absorption for this model. 
Thus, we adopt the partial wave method to calculate the phase shift at the low energy limit. 
{The spherically symmetrical noncommutative black hole with global monopole is described by the following line element}
\begin{eqnarray}
\label{metrncm}
ds^2=F(r)dt^2-\frac{dr^2}{F(r)}-r^2\left(d\vartheta^2 + \sin^2\vartheta d\phi^2 \right).
\end{eqnarray}
The function $ F(r) $ is given by
\begin{eqnarray}
F(r) &=&1-8\pi\eta^2 -\frac{2{\cal M}_{\theta}}{r} ,
\\
&=&1-8\pi\eta^2 -\frac{2M}{r} +\frac{8M\sqrt{\theta}}{\sqrt{\pi}r^2}.
\end{eqnarray}
Here the event horizon, $ r_{\eta} $, and the Cauchy horizon, $ r_{\theta} $, read 
\begin{eqnarray}
\label{rn}
r_{\eta}& =& \frac{2M}{(1-8\pi\eta^2)}-4\sqrt{\frac{\theta}{\pi}}+\cdots
=\frac{2M}{(1-8\pi\eta^2)}\left[ 1-\frac{2(1-8\pi\eta^2)}{M}\sqrt{\frac{\theta}{\pi}}\right],
\\ 
r_{\theta}& =& 4\sqrt{\frac{\theta}{\pi}}+\cdots.
\label{rt}
\end{eqnarray}
The mass $M$ can be written in terms of $r_{\eta}$ as follows
\begin{eqnarray}
M = \left(1-8\pi\eta^{2}\right)\left(\frac{r_{\eta}}{2} + 2\sqrt{\frac{\theta}{\pi}}\right).
\end{eqnarray}
The Hawking temperature is given by
\begin{eqnarray}
T_{H} = \dfrac{\left(1-8\pi\eta^{2}\right)}{4\pi r_{\eta}^{2}}\left(r_{\eta} - 4\sqrt{\frac{\theta}{\pi}}\right).
\label{TemH}
\end{eqnarray}
For $\eta=0$, we have the result obtained in~\cite{Anacleto:2020zfh}.

In order to obtain a minimum mass in the final stage of black hole evaporation, we compute the specific heat capacity using the formula
\begin{eqnarray}
C =\left(\dfrac{\partial M}{\partial r_{\eta}}\right)\left(\dfrac{\partial T_{H}}{\partial r_{\eta}}\right)^{-1}
= -2\pi\left(r_{\eta} - 8\sqrt{\theta/\pi} \right) \left(r_{\eta} + 16\sqrt{\theta/\pi}\right).
\end{eqnarray}
Writing in terms of $r_s$ we have
\begin{eqnarray}
C  = \dfrac{-2\pi\left(r_{s} - 12(1-8\pi\eta^{2})\sqrt{\theta/\pi}\right)\left(r_{s} + 12(1-8\pi\eta^{2})\sqrt{\theta/\pi}\right)}{(1-8\pi\eta^{2})^{2}}.
\end{eqnarray}
Note that for $r_s=r_{min}=12(1-8\pi\eta^{2})\sqrt{\theta/\pi}$, we have $C=0$ and the black hole becomes a remnant. 
So we have $r_{min}=2M_{min}$ and from there we find the following minimum mass
$M_{min}=6(1-8\pi\eta^{2})\sqrt{\theta/\pi}$.

\begin{figure}[!htb]
 \centering
 \includegraphics[scale=0.4]{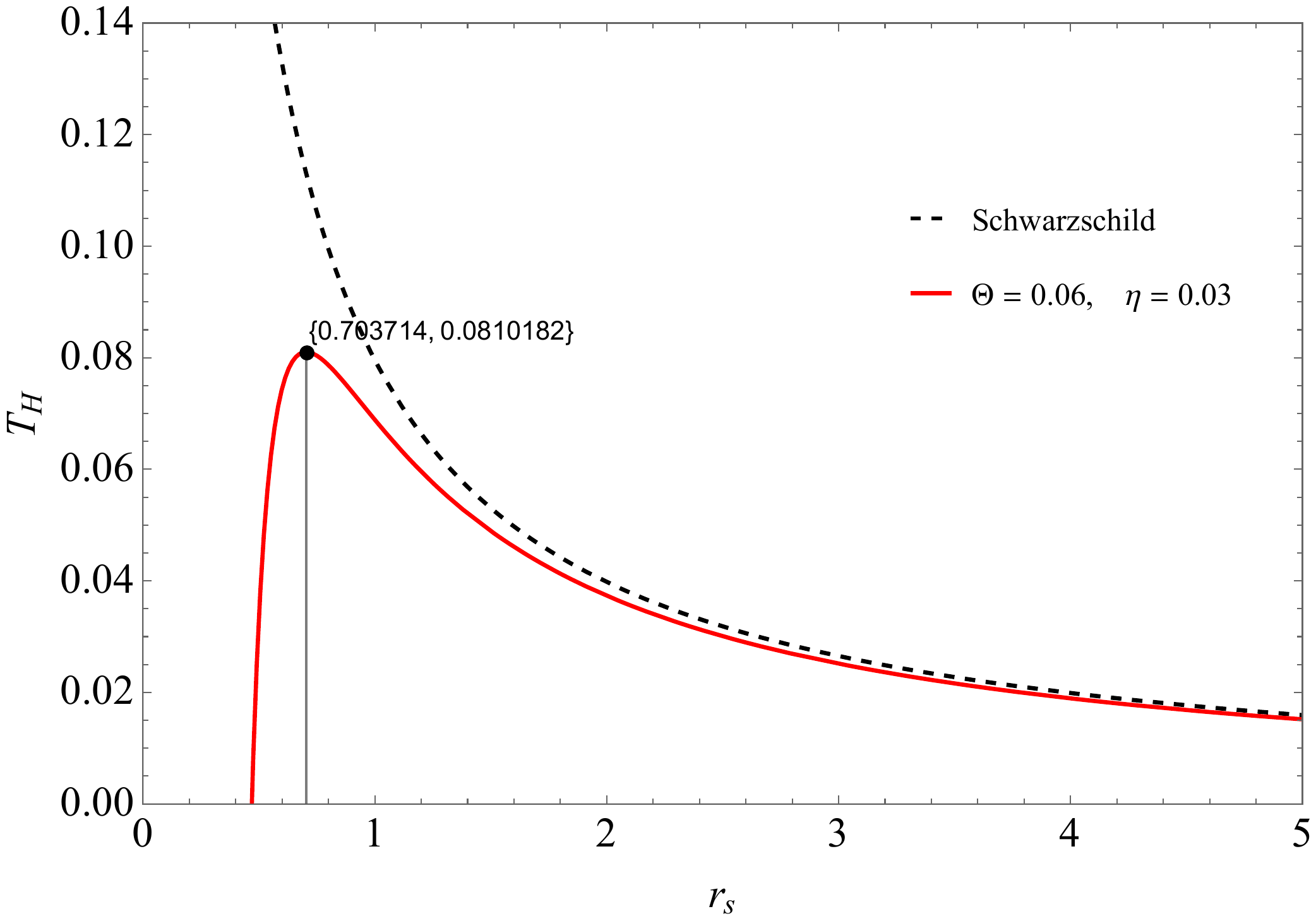}
  \caption{\footnotesize{We have the Hawking temperature for the non-commutative case with monopoles compared to the usual Schwarzschild case. }}
\label{figTmp}
\end{figure}

Therefore, expressing equation \eqref{TemH} in terms of $r_s$, we have
\begin{eqnarray}
T_{H} = \dfrac{\left(1-8\pi\eta^{2}\right)^{2}\left(r_{s} - 8(1-8\pi\eta^{2})\sqrt{\theta/\pi}\right)}{4\pi \left( r_{s} - 4(1-8\pi\eta^{2})\sqrt{\theta/\pi}\right)^{2}}.
\label{TemH2}
\end{eqnarray}
Now substituting the radius $ r_{min} $ into \eqref{TemH2} we find the following maximum temperature of the black hole remnant
\begin{eqnarray}
T_{Hmax} = \dfrac{(1-8\pi\eta^{2})}{64\pi\sqrt{\theta/\pi}}.
\end{eqnarray}
In Fig. \ref{figTmp} we have the graph of temperature as a function of $r_{s}$ for $\Theta=0.06$ and $\eta=0.03$, where $ \Theta=\sqrt{\theta}/(M\sqrt{\pi}) $. For a minimum radius $r_{min} = 0.70371$ and a minimum mass $M_{min}=0.3519$, then the value for the maximum temperature is $T_{Hmax} = 0.08101$. The maximum temperature value corresponds precisely to the peak of the graph.

\subsection{Differential Scattering Cross Section and Absorption}
Now, we start our analysis of the massless scalar field equation to examine the scattered wave in the background (\ref{metrncm}),  given by 
\begin{eqnarray}
\dfrac{1}{\sqrt{-g}}\partial_{\mu}\Big(\sqrt{-g}g^{\mu\nu}\partial_{\nu}\Psi\Big)=0.
\label{eqkg}
\end{eqnarray}
Next, we use the following separation of variables in the  Klein-Gordon equation (\ref{eqkg})
\begin{eqnarray}
\Psi_{\omega l m}({\bf r},t)=\frac{{\cal R}_{\omega l}(r)}{r}Y_{lm}(\vartheta,\phi)e^{-i\omega t},
\end{eqnarray}
being $Y_{lm}(\vartheta,\phi)  $ the spherical harmonics and $ \omega $ the frequency.

Hence, we find the radial equation for $ {\cal R}_{\omega l}(r) $  as follows 
\begin{eqnarray}
\label{eqrad}
F(r)\frac{d}{dr}\left(F(r)\dfrac{d{\cal R}_{\omega l}(r)}{dr} \right) +\left[ \omega^2 -V_{eff} \right]{\cal R}_{\omega l}(r)=0,
\end{eqnarray}
with
\begin{eqnarray}
V_{eff}=\frac{F(r)}{r}\frac{dF(r)}{dr}+\frac{F(r)l(l+1)}{r^2},
\label{efp}
\end{eqnarray}
the effective potential.

Next, we introduce a new radial function, $ \chi(r)=F^{1/2}(r){\cal R}(r) $, such that
\begin{eqnarray}
\label{eqradpsi}
\frac{d^2\chi(r)}{dr^2}+V(r) \chi(r) = 0,
\end{eqnarray}
being $ V(r) $ the potential, given by
\begin{eqnarray}
\label{poteff}
V(r)=\frac{[F'(r)]^2}{4 F^2(r)} - \dfrac{F''(r)}{2F(r)} + \dfrac{\omega^2}{F^2(r)} - 
\frac{V_{eff}}{F^2(r)}.
\end{eqnarray}
At this point, by performing a power series expansion in $1/r$ in the above equation, we find
\begin{eqnarray}
\label{eqrqm}
\frac{d^2\chi(r)}{dr^2}+\left[\frac{\omega^2}{(1-8\pi\eta^2)^2}+ {\cal U}_{eff}(r)\right] \chi(r) = 0,
\end{eqnarray}
The effective potential is now given by
\begin{eqnarray}
\label{pot1}
{\cal U}_{eff}(r)= \frac{4M{\omega}^2}{(1-8\pi\eta^2)^3r}+\frac{12\ell^2}{r^2}
+\frac{32M{\omega}^2}{(1-8\pi\eta^2)^4r^2}\sqrt{\frac{\theta}{\pi}}+\cdots,
\end{eqnarray}
Here, due to the modification of the term $1/r^2$ in the effective potential~\cite{Anacleto:2017kmg,Anacleto:2019tdj}, we define
\begin{eqnarray}
\label{ell}
\ell^2&\equiv&-\frac{(l^2+l)}{12(1-8\pi\eta^2)}+\frac{r_{\eta}^2{\omega}^2}{4(1-8\pi\eta^2)^2}.
\end{eqnarray}
Note that the suitable asymptotic behavior for the effective potential is satisfied, i.e., $ {\cal U}_{eff}(r) \rightarrow 0 $ when $r\rightarrow\infty$.
Next, we can find the phase shift analytically, at the low frequency limit, by means of the following 
Ansatz~\cite{Anacleto:2017kmg,Anacleto:2019tdj}
\begin{eqnarray}
\label{formapprox}
\delta_l\approx 2(l-\ell)=2\left(l - \sqrt{-\frac{(l^2+l)}{12(1-8\pi\eta^2)}
+\frac{r^2_{\eta}{\omega}^2}{4(1-8\pi\eta^2)^2}}\right). 
\end{eqnarray}
Hence, the phase shift $ \delta_{l} $  for $ l\rightarrow 0 $ becomes
\begin{eqnarray}
\label{phase2}
\delta_l=\delta_0 + \delta_{l\geq 1}, \quad 
\delta_0=-\frac{{\omega}r_{\eta}}{(1-8\pi\eta^2)},
\quad \delta_{l\geq 1}=0 .
\end{eqnarray}
{Furthermore, with determined  $\delta_l$ we can calculate the differential scattering cross section 
using the formula~\cite{Anacleto:2017kmg,Yennie1954,Cotaescu:2014jca}
\begin{eqnarray}
\label{scattdif}
\dfrac{d\sigma}{d\Omega}
=\Big| \frac{1}{2i{{\omega}}}\sum_{l=0}^{\infty}
a^1_{l}\frac{P_{l}(\cos\vartheta)}{1-\cos\vartheta}\Big|^2,
\end{eqnarray}
where $ P_l(\cos\vartheta) $ are the Legendre polynomials, being
\begin{eqnarray}
\label{relrec}
a^1_{l}=a^0_{l}-\frac{l+1}{2l+3}a^0_{l+1}-\frac{l}{2l-1}a^0_{l-1},
\end{eqnarray}
and
\begin{eqnarray}
a^0_l=(2l+1)\left(e^{2i\delta_l} -1 \right).
\label{al0}
\end{eqnarray}
Therefore, from expression (\ref{relrec}) for $l=0$, we have
\begin{eqnarray}
a^1_0=a^0_0 -\frac{1}{3}a^0_1,
\end{eqnarray}
and from relation (\ref{al0}), we have as a result:
\begin{eqnarray}
a^0_0=e^{2i\delta_0} -1,
\end{eqnarray}
and
\begin{eqnarray}
a^0_1=3\left(e^{2i\delta_1} -1\right)=0,
\end{eqnarray}
because $ \delta_{l\geq 1}=0 $. Then the equation for the differential scattering cross section (\ref{scattdif}) can be now  written as follows
\begin{eqnarray}
\label{espalh}
\dfrac{d\sigma}{d\Omega}=\Big| \frac{1}{2i{{\omega}}}\sum_{l=0}^{\infty}
a^0_{l}\frac{P_{l}(\cos\vartheta)}{1-\cos\vartheta}\Big|^2=\Big| \frac{1}{2i{{\omega}}}\sum_{l=0}^{\infty}(2l+1)\left(e^{2i\delta_l} -1 \right)
\frac{P_{l}(\cos\vartheta)}{1-\cos\vartheta}\Big|^2,
\end{eqnarray}
and that at the small angle limit equation (\ref{espalh}) takes the form
\begin{eqnarray}
\label{espalh2}
\frac{d\sigma}{d\Omega}&=&\frac{4}{{\omega}^2\vartheta^4}\Big|\sum_{l=0}^{\infty}(2l+1)
e^{i\delta_l}\sin(\delta_{l})
{P_{l}(\cos\vartheta)}\Big|^2,
\\
&=&\frac{4}{{\omega}^2\vartheta^4}\Big|e^{i\delta_0}\sin(\delta_0)P_0(\cos\vartheta)+\sum_{l=1}^{\infty}(2l+1)e^{i\delta_{l\geq 1}}\sin(\delta_{l\geq 1})
{P_{l}(\cos\vartheta)}\Big|^2.
\label{espalh3}
\end{eqnarray}
Moreover, at the low frequency limit, by using the result of (\ref{phase2}), we find the following result for 
the differential scattering cross section }
\begin{eqnarray}
\label{diffcsect}
\frac{d\sigma}{d\Omega}\Big |^{\mathrm{l f}}_{{\omega}\rightarrow 0}
=\frac{4}{{\omega}^2\vartheta^4}\delta^2_0=\frac{4}{\vartheta^4}\frac{r_{\eta}^2}{(1-8\pi\eta^2)^2}
=\frac{16M^2}{(1-8\pi\eta^2)^4\vartheta^4}\left( 1-\frac{2(1-8\pi\eta^2)}{M}\sqrt{\frac{\theta}{\pi}}\right)^2+\cdots.
\end{eqnarray}
For $ \eta=0 $ and $ \theta=0 $ we obtain the result for the Schwarzschild black hole case.
Now, an interesting result for the differential scattering cross section emerges when we take $ M\rightarrow 0$ and thus find a non-zero result given by
\begin{eqnarray}
\label{diffcsect2}
\frac{d\sigma}{d\Omega}\Big |^{\mathrm{l f}}_{M\rightarrow 0}
\approx\frac{64\theta}{\vartheta^4 \pi(1-8\pi\eta^2)^2}=\frac{16 M^2_{min}}{9\vartheta^4 (1-8\pi\eta^2)^4},
\end{eqnarray}
where $ M_{min}=6(1-8\pi\eta^{2})\sqrt{\theta/\pi} $ is the minimal mass~\cite{Anacleto:2020zfh}. 
Note that the differential scattering cross section is increased by the monopole effect.

\begin{figure}[!htb]
 \centering
 \includegraphics[scale=0.45]{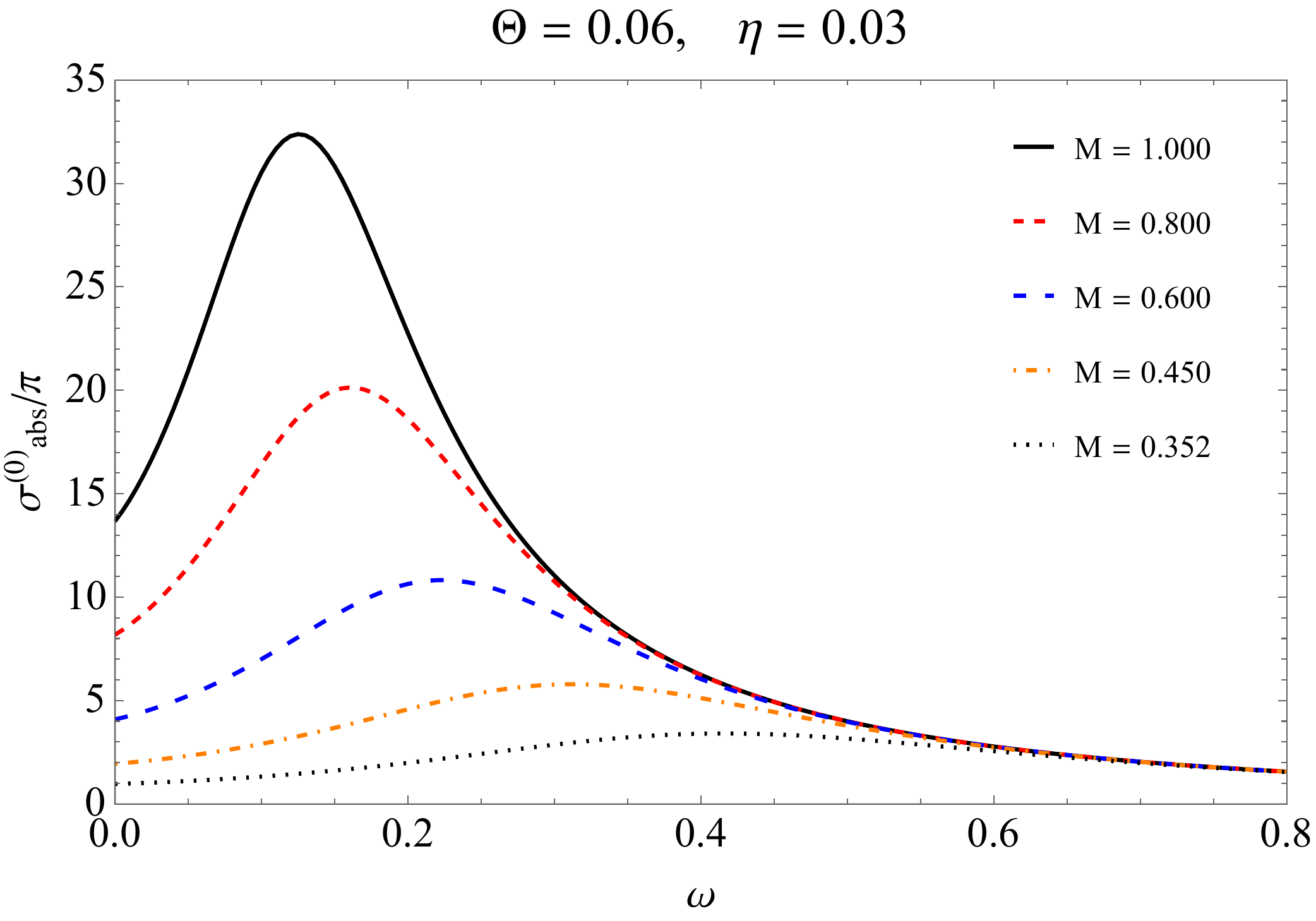}
  \caption{\footnotesize{Here we see the behavior of the absorption cross section for $l = 0$ when varying $M$ from 1 to the minimum mass value, corresponding to the values of the parameters $\Theta$ and $\eta$.}} 
\label{absM0235}
\end{figure}

At this point, we will compute the absorption cross section for a noncommutative black hole with a global monopole, at the low frequency  limit, which can be obtained as follows:
\begin{eqnarray}
\label{abdf}
\sigma_{abs}&=&\frac{\pi}{{\omega}^2}\sum_{l=0}^{\infty}(2l+1)\Big(1-\big|e^{2i\delta_l}\big|^2\Big)
=\frac{4\pi}{{\omega}^2}\sum_{l=0}^{\infty}(2l+1)\sin^2(\delta_{l}),
\\
&=&\frac{4\pi}{{\omega}^2}\left[\sin^2(\delta_{0})+\sum_{l=1}^{\infty}(2l+1)\sin^2(\delta_{l\geq 1})\right].
\end{eqnarray}
Now, by taking the limit $ {\omega}\rightarrow 0 $  with $ \delta_l $ given by (\ref{phase2}) the absorption reads 
\begin{eqnarray}
\label{abs1}
\sigma_{abs}^{\mathrm{l f}}&=&\frac{4\pi\delta_0^2}{{\omega}^2}=\frac{4\pi r^2_{\eta}}{(1-8\pi\eta^2)^2}
= \frac{16\pi M^2}{(1-8\pi\eta^2)^4}\left( 1-\frac{2(1-8\pi\eta^2)}{M}\sqrt{\frac{\theta}{\pi}}\right)^2+\cdots.
\label{absma}
\end{eqnarray}
We can recover the result for the absorption of the Schwarzschild black hole in the absence of noncommutativity and global monopole by making $\eta = \theta = 0$.
Furthermore, an interesting result for the absorption arises when the mass parameter goes to zero and in this case 
we obtain a non-zero result given by
\begin{eqnarray}
\label{abs1a}
\sigma_{abs}^{\mathrm{l f}}
&\approx &\frac{64\theta}{(1-8\pi\eta^2)^2}=\frac{16\pi M^2_{min}}{9(1-8\pi\eta^2)^4}.
\end{eqnarray}
In Fig.~\ref{absM0235} we show the behavior of the absorbing cross section for $l = 0$ when varying $M$ from 1 to the minimum mass value.
Note that by varying the $\eta$ parameter, the absorption has its value increased.
Here, it is worth mentioning that the results obtained in (\ref{diffcsect2}) and (\ref{abs1a}) are not present in the usual case of the Schwarzschild black hole when the mass parameter approaches zero. 

Besides, for absorption we can express the result in terms of the area of the event horizon of the noncommutative black hole with global monopole, so we have
\begin{eqnarray}
\label{abs2}
\sigma_{abs}^{\mathrm{l f}}
=\frac{4\pi{r_{\eta}^2}}{(1-8\pi\eta^2)^2}
\approx{\cal A}.
\end{eqnarray}
We can also express Eq.~\eqref{abs1a} in terms of a minimum area as follows
\begin{eqnarray}
\label{abs1amin}
\sigma_{abs,\theta}^{\mathrm{l f}}
=\frac{4\pi r^2_{min}}{9(1-8\pi\eta^2)^4}=\frac{A_{min}}{9(1-8\pi\eta^2)^4},
\end{eqnarray}
where $ A_{min}=4\pi r^2_{min}$ is a minimal area.
Therefore, we verified that at the limit of $ M $ going to zero, the black hole becomes a remnant of black hole 
and for $ \eta = 0 $ the absorption is equal to a minimal area due to the effect of noncommutativity.

{Now, we will calculate the absorption at the high frequency limit. For this we rewrite the potential (\ref{pot1}) as follows
\begin{eqnarray}
{\cal U}_{eff}(r)= \frac{4M{\omega}^2}{(1-8\pi\eta^2)^3r}+\frac{\bar{\ell}^{\,2}}{(1-8\pi\eta^2)r^2}
+\frac{48M{\omega}^2}{(1-8\pi\eta^2)^4r^2}\sqrt{\frac{\theta}{\pi}}+\cdots,
\end{eqnarray}
where
\begin{eqnarray}
\bar{\ell}^{\,2}\equiv&(l^2+l)-\dfrac{3\left(r_{\eta}^2+16M\sqrt{\theta/9\pi}\right){\omega}^2}{(1-8\pi\eta^2)}.
\end{eqnarray}
In this case, to determine the phase shift we apply the following Ansatz
\begin{eqnarray}
\delta_l\approx\dfrac{3}{4}\sqrt{l-\bar{\ell}\,}.
\end{eqnarray}
For $ l\rightarrow\infty $ the phase shift is given by
\begin{eqnarray}
\delta_l\approx\dfrac{3}{4}\sqrt{\dfrac{3}{2}\dfrac{\left(r_{\eta}^2+16M\sqrt{\theta/9\pi}\right){\omega}^2}
{(1-8\pi\eta^2) l}}.
\end{eqnarray}
Therefore, at the limit of $ l\rightarrow\infty $ , the absorption is given by
\begin{eqnarray}
\sigma_{abs}=\frac{4\pi}{{\omega}^2}\sum_{l=0}^{\infty}2l\delta_{l}^2
=\frac{4\pi}{{\omega}^2}\sum_{l=0}^{\infty}2l\dfrac{27}{16}\dfrac{\left(r_{\eta}^2+16M\sqrt{\theta/9\pi}\right){\omega}^2}{2l(1-8\pi\eta^2) },
\end{eqnarray}
such that we obtain the following result for absorption at the high frequency limit 
\begin{eqnarray}
\sigma_{abs}^{hf}
=\dfrac{27\pi}{4}\dfrac{\left(r_{\eta}^2+16M\sqrt{\theta/9\pi}\right)}{(1-8\pi\eta^2)}=\frac{27\pi M^2}{(1-8\pi\eta^2)^3}\left( 1-\frac{8(1-8\pi\eta^2)}{3M}\sqrt{\frac{\theta}{\pi}}\right).
\end{eqnarray}
Here we have successfully obtained absorption at the high frequency limit by the partial wave method.
Note that for $ \eta=\theta=0 $, we recover the absorption result for the Schwarzschild black hole at the high frequency limit obtained usually by the null geodesic method, $ \sigma_{abs}^{hf}=27\pi M^2 $.

For $ \eta=0 $ and $ \theta\neq 0 $, we find
\begin{eqnarray}
\sigma_{abs}^{hf}=27\pi M^2 - 72\pi M\sqrt{\frac{\theta}{\pi}}+\cdots,
\end{eqnarray}
showing that the absorption is reduced when we vary $\theta$.

For $ \eta\neq 0 $ and $ \theta= 0 $, we obtain
\begin{eqnarray}
\sigma_{abs}^{hf}=27\pi M^2(1+24\pi\eta^2 + \cdots).
\end{eqnarray}
In this case the absorption is increased when we vary $\eta$,
which coincides with the results obtained by the null geodesic method.
}
\subsection{Numerical analyses}
In order to investigate the behavior of the radial equation (\ref{eqrad}) in the asymptotic limits we define the tortoise coordinate $ \varrho $ as follows:
\begin{eqnarray}
\frac{d}{d\varrho}=F(r)\frac{d}{dr}, \quad\quad F(r)=1-8\pi\eta^2 -\frac{2M}{r} +\frac{8M\sqrt{\theta}}{\sqrt{\pi}r^2},
\end{eqnarray}
such that 
\begin{eqnarray}
\varrho=r+\frac{r^2_{\theta}}{r_{\theta}-r_{\eta}}\ln\Big|{r-r_{\theta}}\Big|
- \frac{r^2_{\eta}}{r_{\theta}-r_{\eta}}\ln\Big|{r-r_{\eta}}\Big|.
\end{eqnarray}
Thus, the radial equation (\ref{eqrad}) can be written as
\begin{eqnarray}
\label{eqradtor}
\dfrac{d^2{\cal R}_{\omega l}(r)}{d\varrho^2} +\left[ \omega^2 -V_{eff} \right]{\cal R}_{\omega l}(r)=0.
\end{eqnarray}
Note that, when applying the limits $ r \rightarrow r_ {\eta} $ and $ r \rightarrow \infty $, the effective potential defined in (\ref{efp}) tends to zero. 
Consequently, we can examine the scattering problem by exploring the radial solution at these asymptotic limits.
In this way, we have the following boundary conditions:
\begin{eqnarray}
R_{\omega l}(r)\approx \Big\{
\begin{array}{cc}
A_{in}e^{-iw\varrho} + A_{out}e^{iw\varrho}, &\quad \varrho\rightarrow +\infty\quad (r\rightarrow \infty), \\ 
A_{tr}e^{-iw\varrho}, &\quad \varrho\rightarrow -\infty\quad (r\rightarrow r_{\eta}),
\end{array} 
\end{eqnarray}
with $ |A_{in}|^2=|A_{out}|^2 +|A_{tr}|^2 $. 
Next, we determine the absorption by applying equation (\ref{abdf}) with the phase shift given by
\begin{eqnarray}
e^{2i\delta_l}=(-1)^{l+1}\frac{A_{out}}{A_{in}}.
\end{eqnarray}

Here we present the numerical results that were obtained by numerically solving the radial equation (\ref{eqrad}). 
For this purpose we have adopted the numerical procedure performed in~\cite{Dolan:2012yc}. 

In the table~\ref{tab1} we show the comparison between the analytical and numerical results for $\omega\rightarrow 0$ and $l=0$. 

\begin{figure}[!htb]
 \centering
 \subfigure[]{\includegraphics[scale=0.35]{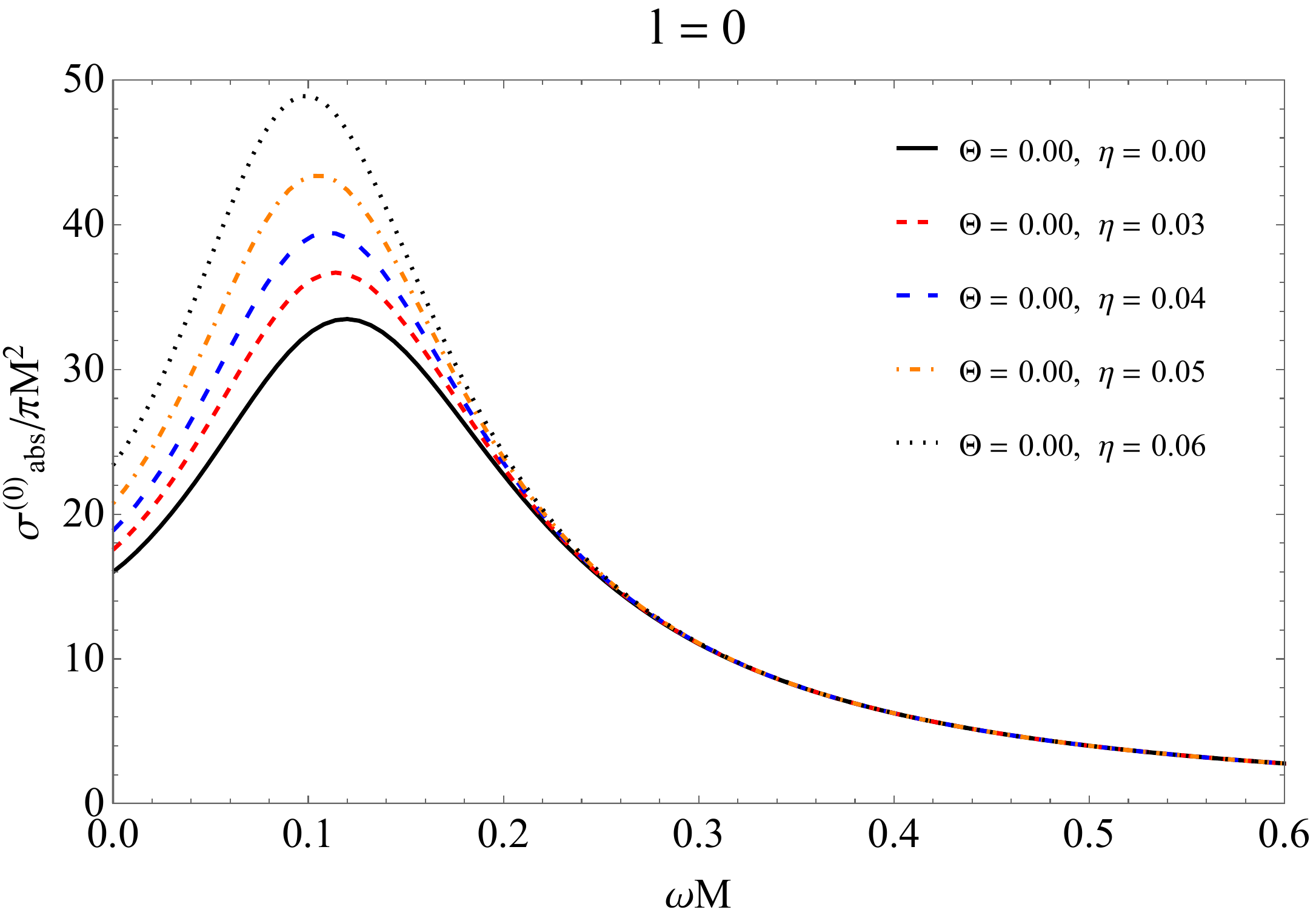}\label{fig1}}
 \qquad
 \subfigure[]{\includegraphics[scale=0.35]{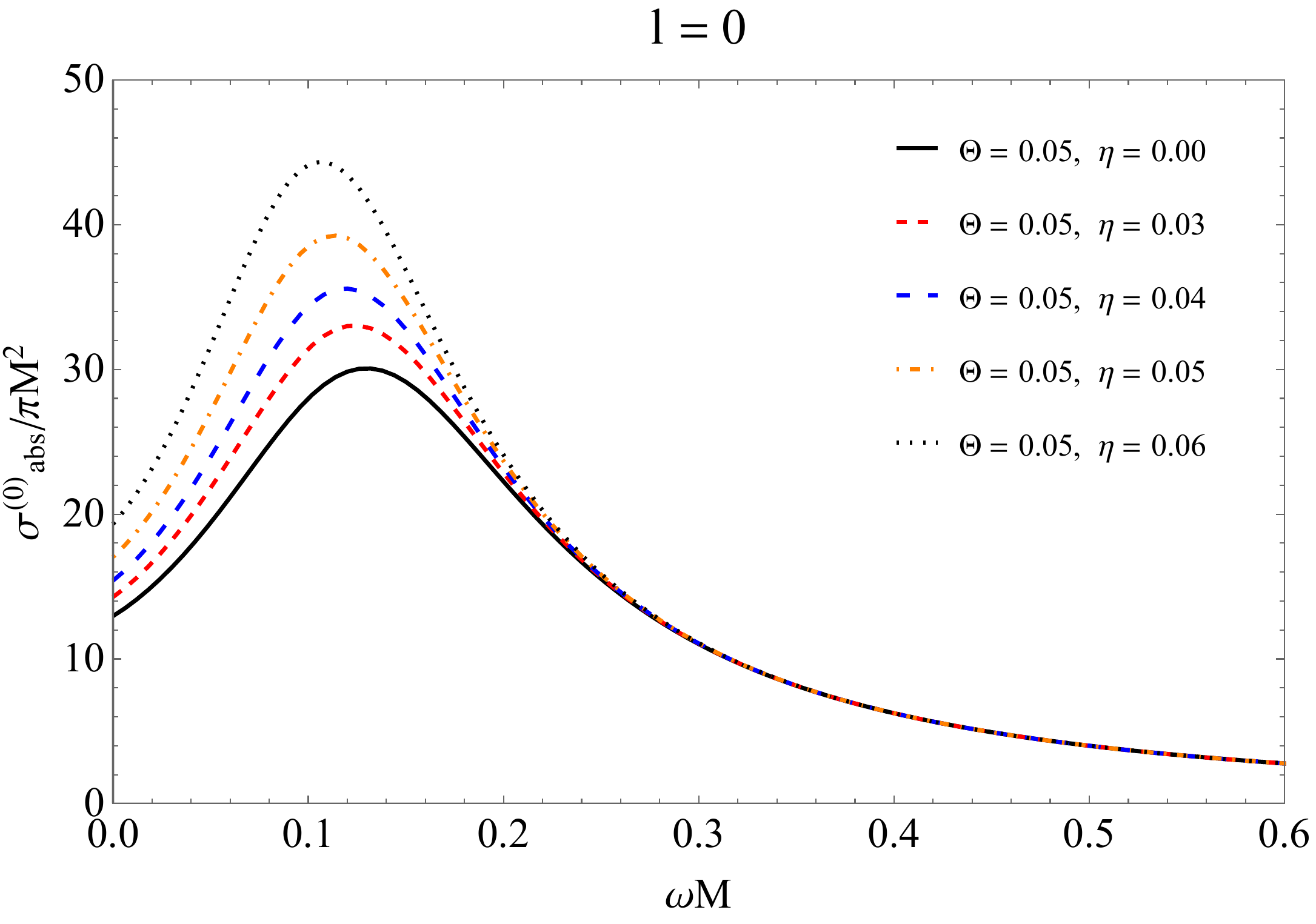}\label{fig2}}
 \qquad
 \subfigure[]{\includegraphics[scale=0.35]{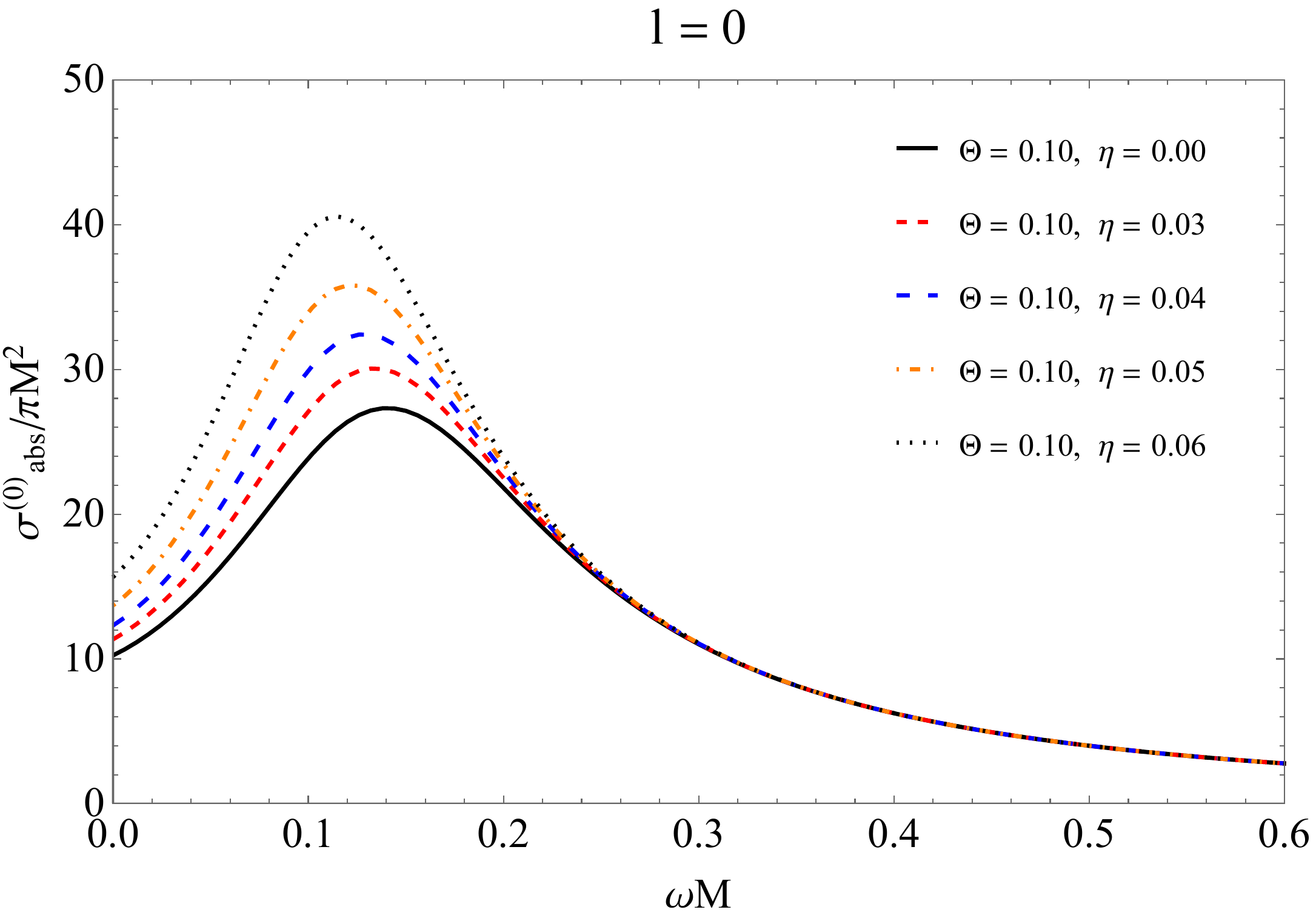}\label{fig3}}
 \qquad
 \subfigure[]{\includegraphics[scale=0.35]{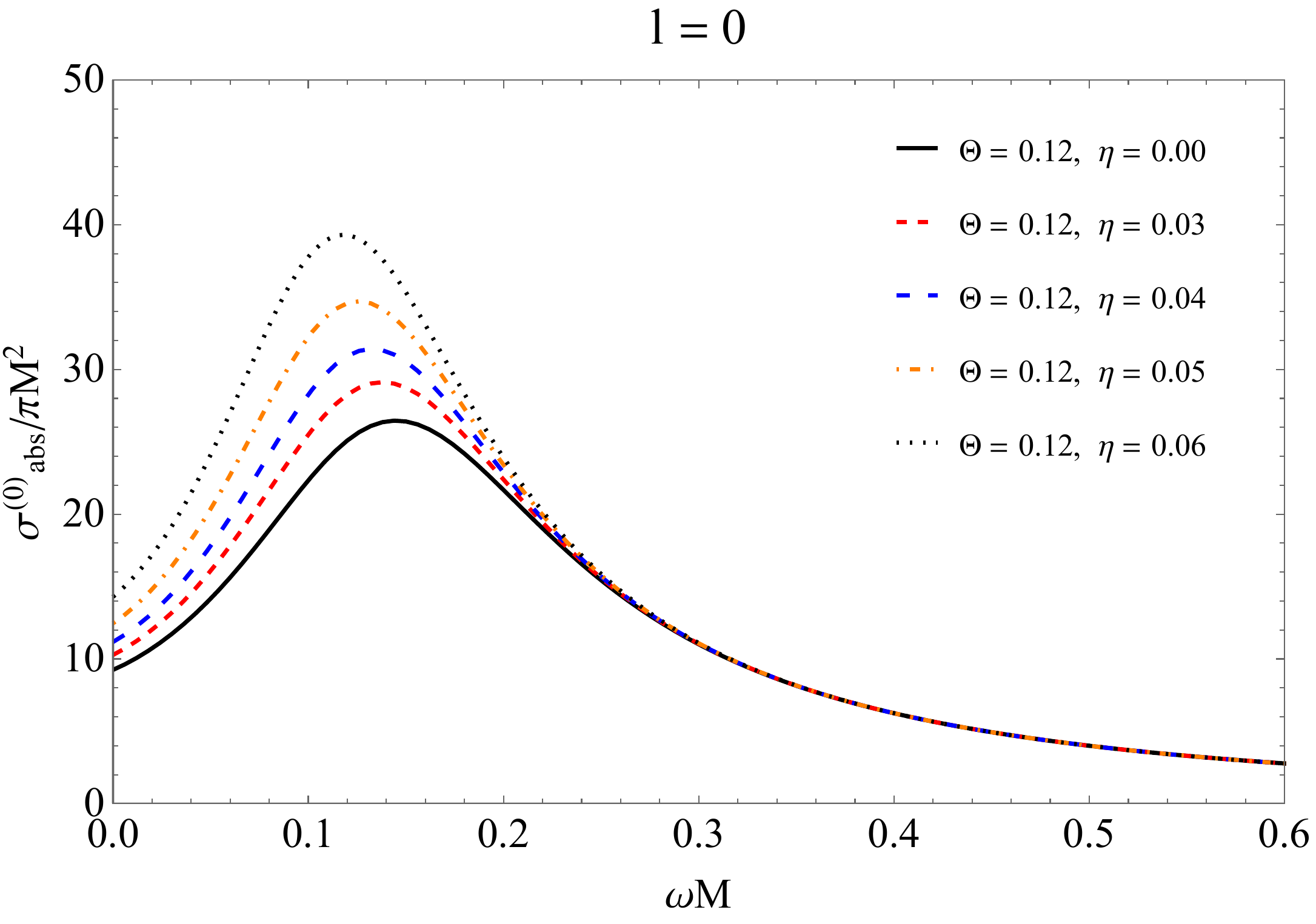}\label{fig4}}
 \qquad
 \subfigure[]{\includegraphics[scale=0.35]{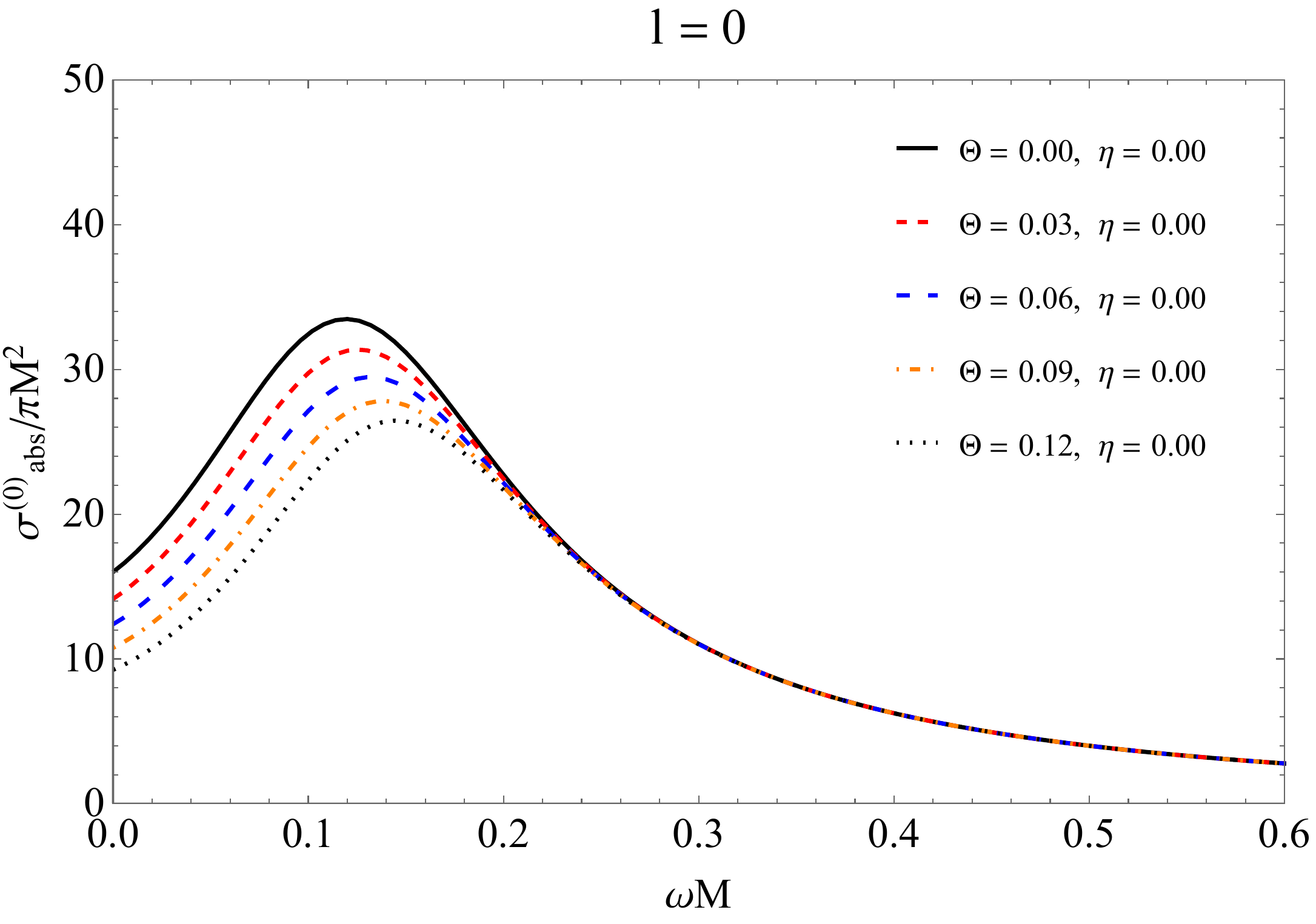}\label{fig5}}
 \qquad
 \subfigure[]{\includegraphics[scale=0.35]{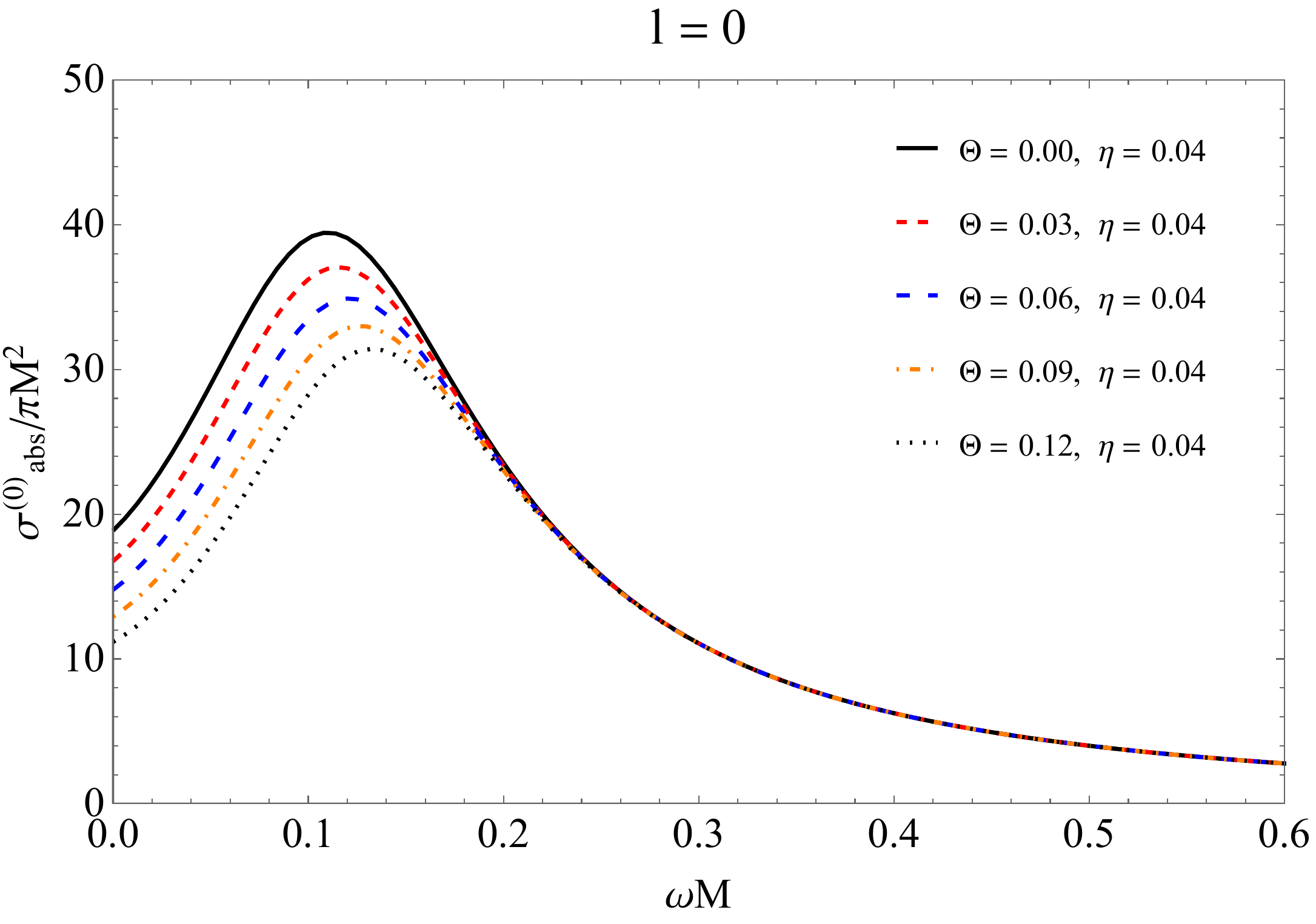}\label{fig6}}
 \\
 \caption{Absorption cross section for $l=0$.} 
 \label{fig}
\end{figure}

In Fig.~\ref{fig}, we plot the partial absorption cross section for the $ l=0 $ mode for different values of $ \Theta=\sqrt{\theta}/(M\sqrt{\pi}) $ and $ \eta $. 
We can see by comparing the curves (Fig.~\ref{fig1}, Fig.~\ref{fig2}, Fig.~\ref{fig3} and Fig.~\ref{fig4})
for different values of $ \Theta=\sqrt{\theta}/(M\sqrt{\pi}) $ and $ \eta $ that the partial absorption is increased 
in relation to the Schwarzschild black hole when we vary $\eta$ with fixed $\Theta$.
Now, in Fig.~\ref{fig5} and Fig.~\ref{fig6} when vary $ \Theta $ with fixed $ \eta $ the partial absorption
is decreased due to the noncommutative effect.  
For $ \Theta=0 $ and $ \eta=0 $ the graph shows the result of the partial absorption for the Schwarzschild black hole. 

\begin{figure}[!htb]
 \centering
 \includegraphics[scale=0.35]{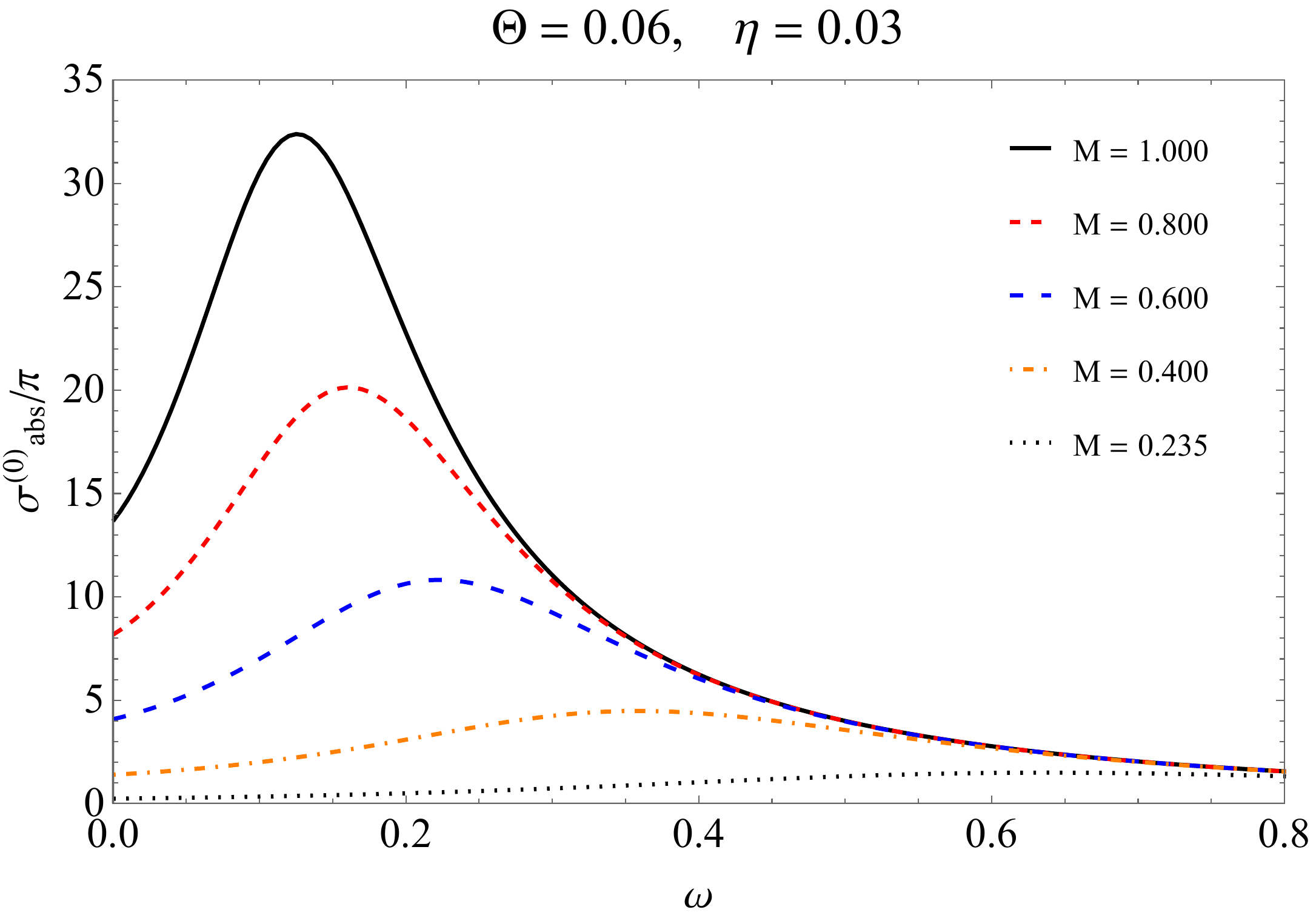}
 \\
  \caption{Absorption cross section for small $ M $.} 
 \label{smallM}
\end{figure}

\begin{figure}[!htb]
 \centering
 \subfigure[]{\includegraphics[scale=0.35]{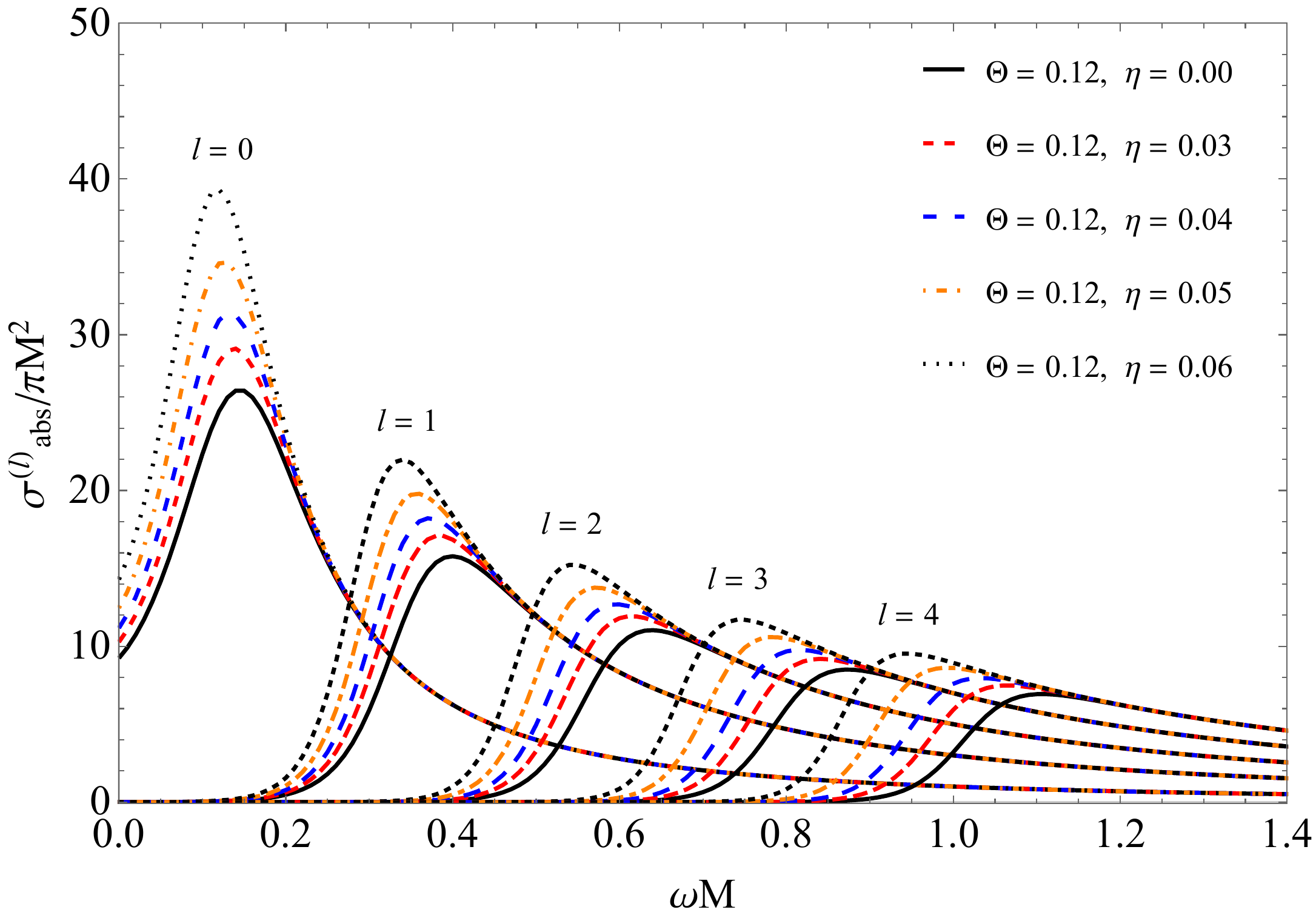}\label{absteta012}}
 \qquad
 \subfigure[]{\includegraphics[scale=0.35]{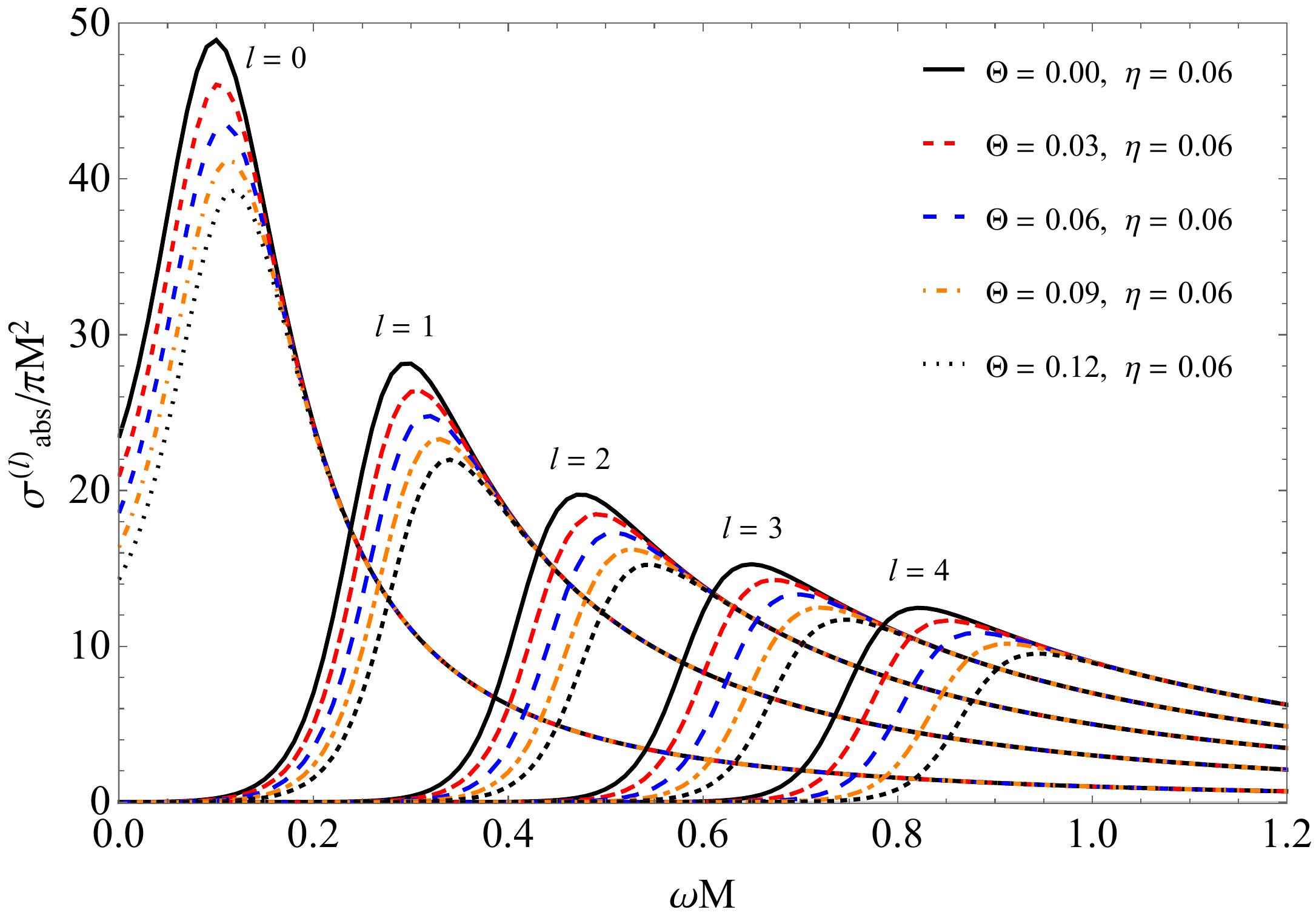}\label{abseta006}}
 \\
  \caption{Absorption cross section for $l=0,1,2,3,4$.} 
\end{figure}

In Fig.~\ref{smallM} we plot the partial absorption for $ l=0 $ mode by setting
the values of $ \Theta=0.06 $ and $ \eta=0.03 $. We can observe that when we reduce the mass value, the absorption amplitude is not null as we can see from equation (\ref{abs1a}). The partial absorption cross section graphs for
$ l=0, 1, 2, 3, 4 $ modes are shown in Fig.~\ref{absteta012}.

\begin{figure}[!htb]
 \centering
 \subfigure[]{\includegraphics[scale=0.35]{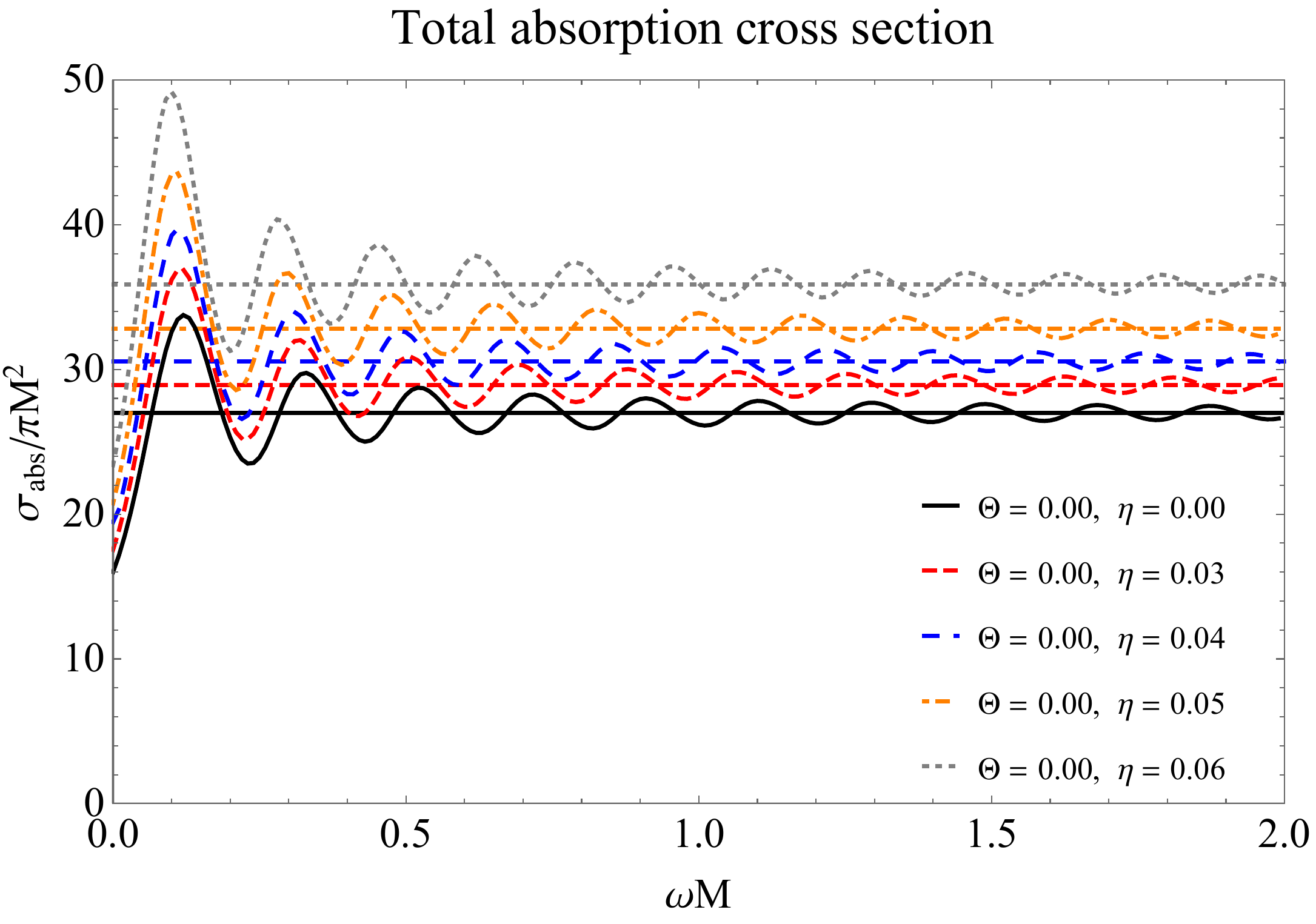}\label{absta}}
 \qquad
 \subfigure[]{\includegraphics[scale=0.35]{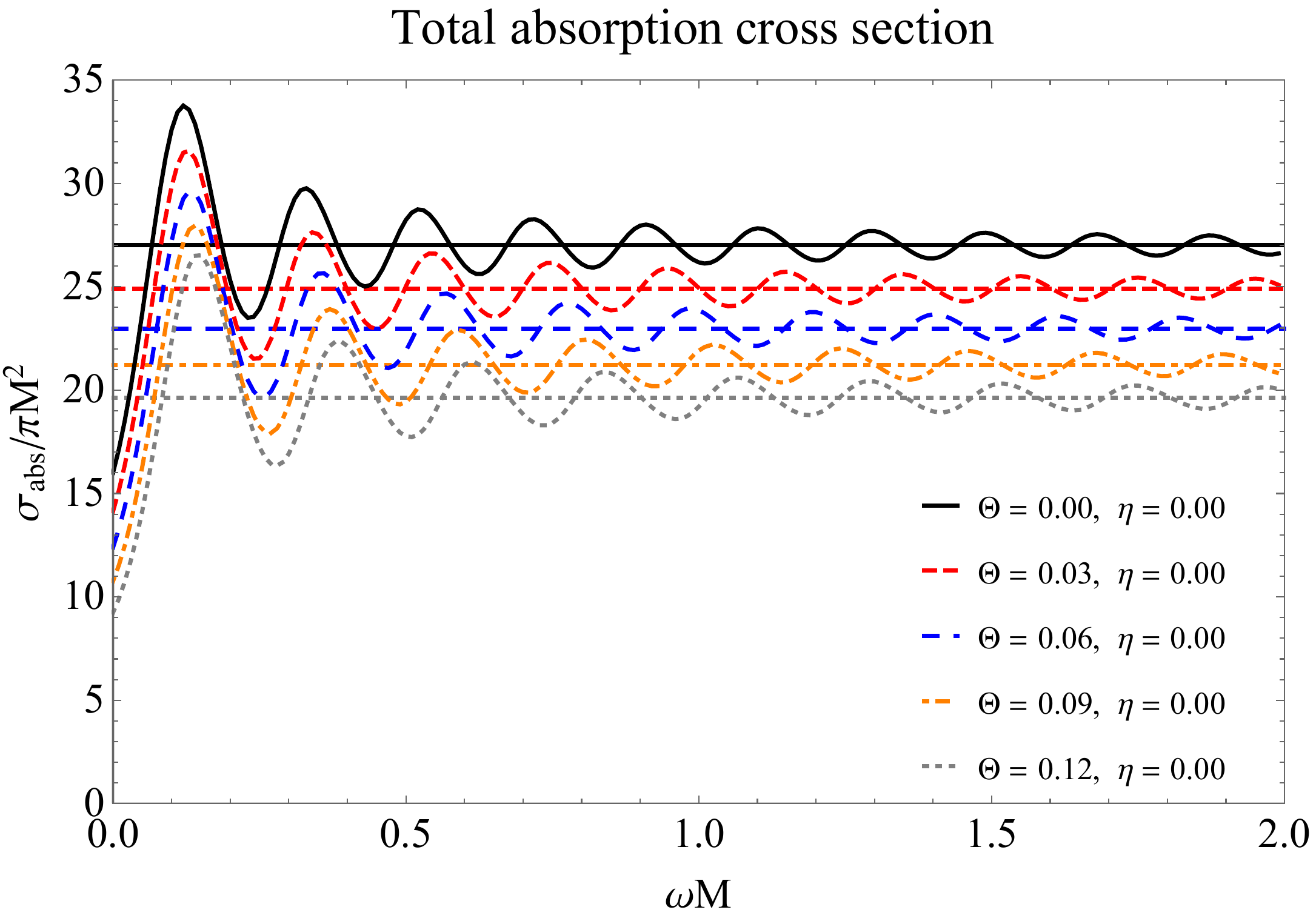}\label{abstb}}
 \qquad
  \subfigure[]{\includegraphics[scale=0.35]{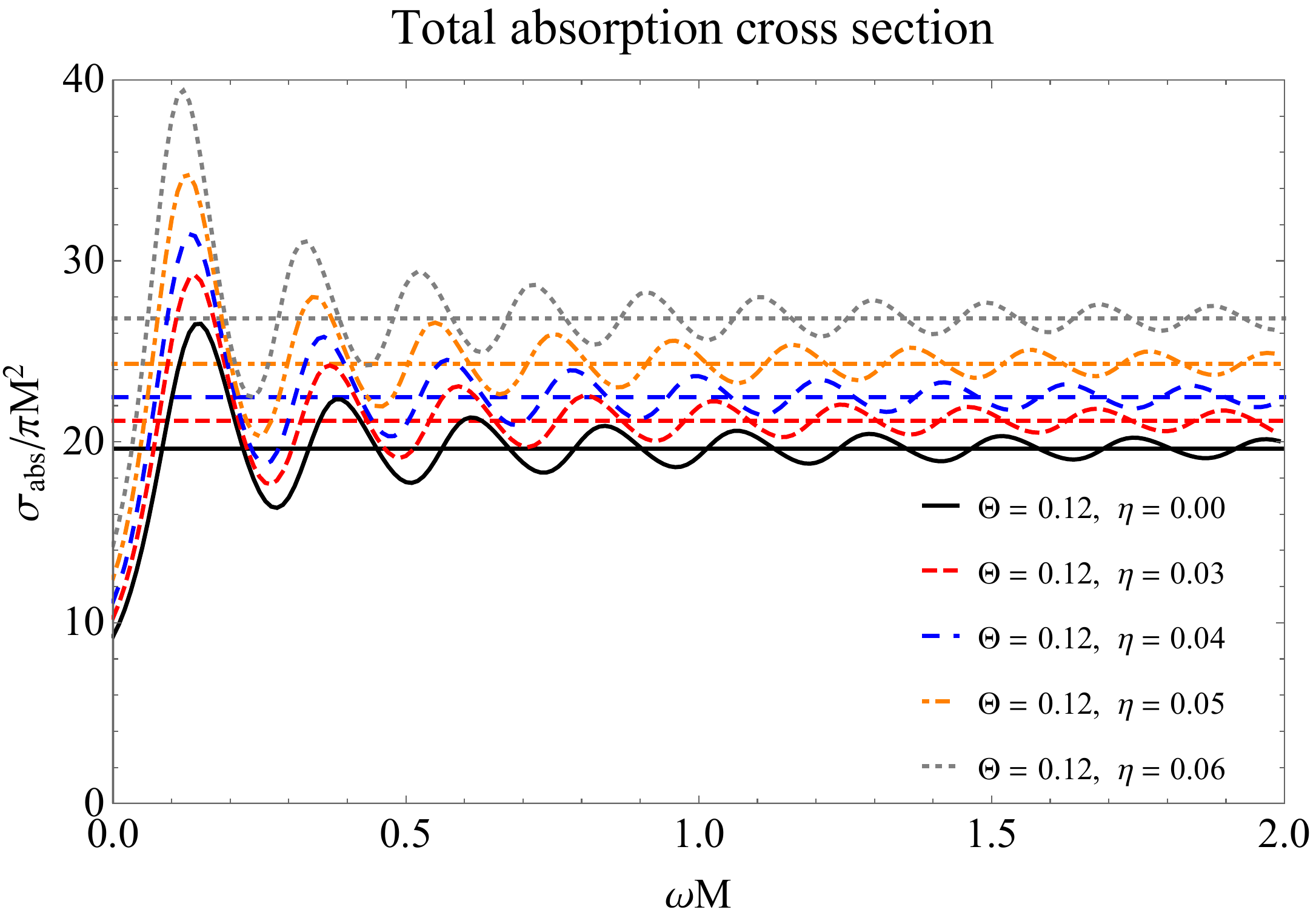}\label{abstc}}
 \qquad
 \subfigure[]{\includegraphics[scale=0.35]{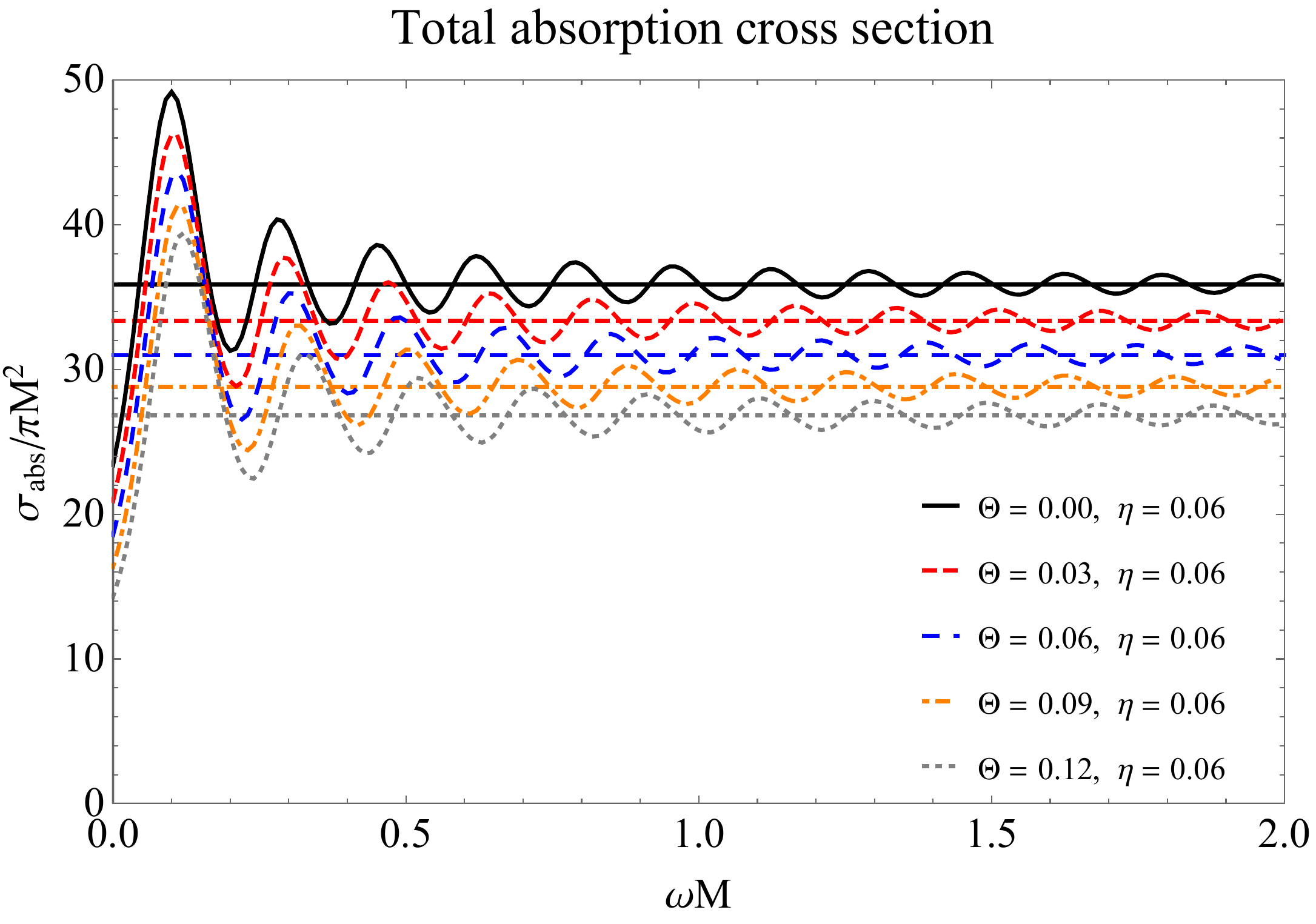}\label{abstd}}
 \caption{Total absorption cross section. The lines represent the values for high frequency absorption, the values obtained by the null geodesics are: (a) $27.0, 28.9, 30.5, 32.8, 35.9$, (b) $ 27.0, 24.9, 23.0, 21.2, 19.6 $, (c) $ 19.6, 21.2, 22.5, 24.3, 26.8 $ and (d) $35.9, 33.4, 31.0, 28.8, 26.8$.} 
 \label{abstotal}
\end{figure}

In Fig.~\ref{abstotal} we present the graphs of the total absorption cross section. 
The results for fixed $\Theta$ and varying $\eta$ values are shown in figures \ref{absta} and \ref{abstc}. 
Graphs for fixed $\eta$ and varying $\Theta$ values are shown in figures \ref{abstb} and \ref{abstd}.
The lines in Fig.~\ref{abstotal} represent the values for high frequency absorption, by applying the null geodesic method presented in the next section.

\begin{figure}[!htb]
 \centering
 \includegraphics[scale=0.4]{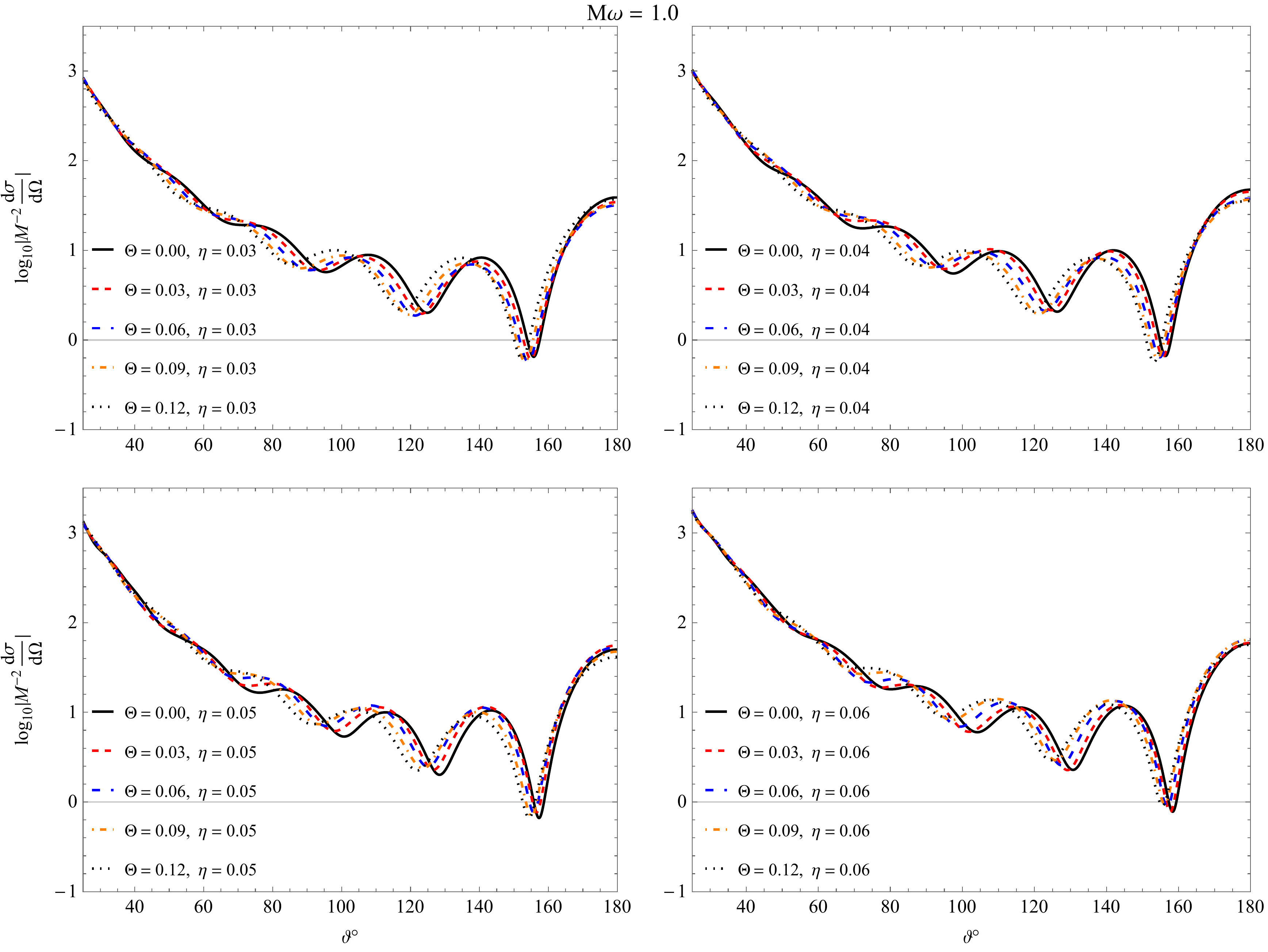} 
 \caption{\footnotesize{Differential scattering cross section for $ M\omega=1.0 $.}} 
 \label{scattMw1}
\end{figure}

\begin{figure}[!htb]
 \centering
 \includegraphics[scale=0.4]{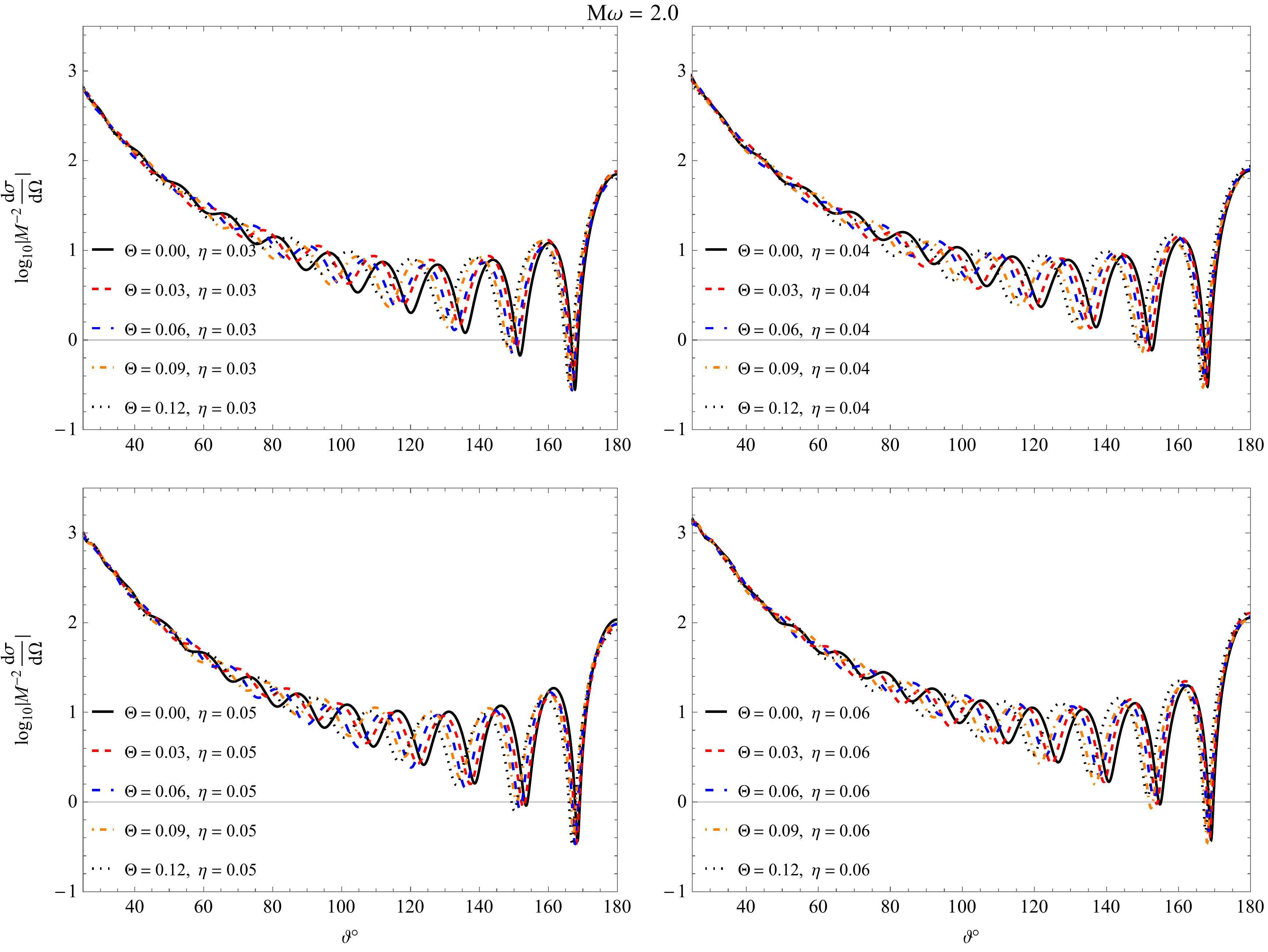} 
 \caption{\footnotesize{Differential scattering cross section for $ M\omega=2.0 $.}} 
 \label{scattMw2}
\end{figure}

Finally in Fig.~\ref{scattMw1} we plot the result of the differential scattering cross section for $ M\omega=1.0 $ and in Fig.~\ref{scattMw2} for $ M\omega=2.0 $. The effect of noncommutativity and the global monopole on
the differential scattering cross section is more significant at large angles.
\begin{table}
\caption{ Analytical and numerical results for $ \omega\rightarrow 0$ and $ l=0 $.}
\begin{tabular}{ |c||c|c||c|c||c|c||c|c|  }
 \hline
  &\multicolumn{2}{c||}{$ \Theta=0.00 $} & \multicolumn{2}{c||}{$ \Theta=0.05 $} & \multicolumn{2}{c||}{$ \Theta=0.10 $}& \multicolumn{2}{c|}{$ \Theta=0.12 $}\\
 \hline
 $\eta$ & Eq.~\eqref{abs1a} & Numerical & Eq.~\eqref{abs1a} & Numerical & Eq.~\eqref{abs1a} & Numerical & Eq.~\eqref{abs1a} & Numerical\\
 \hline
 0.00  & 16.0000 & 15.9988 & 12.9600 &  12.9592 & 10.2400 & 10.2396 & 9.24160 & 9.24105\\
 0.03  & 17.5334 & 17.5324 & 14.2735 &  14.2729 & 11.3486 & 11.3484 & 10.2725 & 10.2723\\
 0.05  & 20.7420 & 20.7400 & 17.0364 &  17.0351 & 13.6952 & 13.6942 & 12.4607 & 12.4598\\
 0.07  & 27.0657 & 27.0603 & 22.5273 &  22.5249 & 18.4051 & 18.4031 & 16.8727 & 16.8703\\
 0.09  & 39.7687 & 39.7637 & 33.6864 &  33.6867 & 28.1086 & 28.1063 & 26.0187 & 26.0171\\
 0.12  & 96.5153 & 96.5062 & 84.5912 &  84.5858 & 73.4531 & 73.4475 & 69.2179 & 69.2136\\
  \hline
\end{tabular}
\label{tab1}
\end{table}

 \section{Null Geodesics and Shadow}
 \label{ng}
In this section, we will apply the null geodesic method to compute the absorption at the high frequency limit and
we will also determine the shadow radius of the noncommutative black hole with global monopole.
Therefore, at this limit, we check the validity of our numerical results.
\subsection{Null Geodesics}
\begin{figure}[!htb]
 \centering
 \subfigure[]{\includegraphics[scale=0.35]{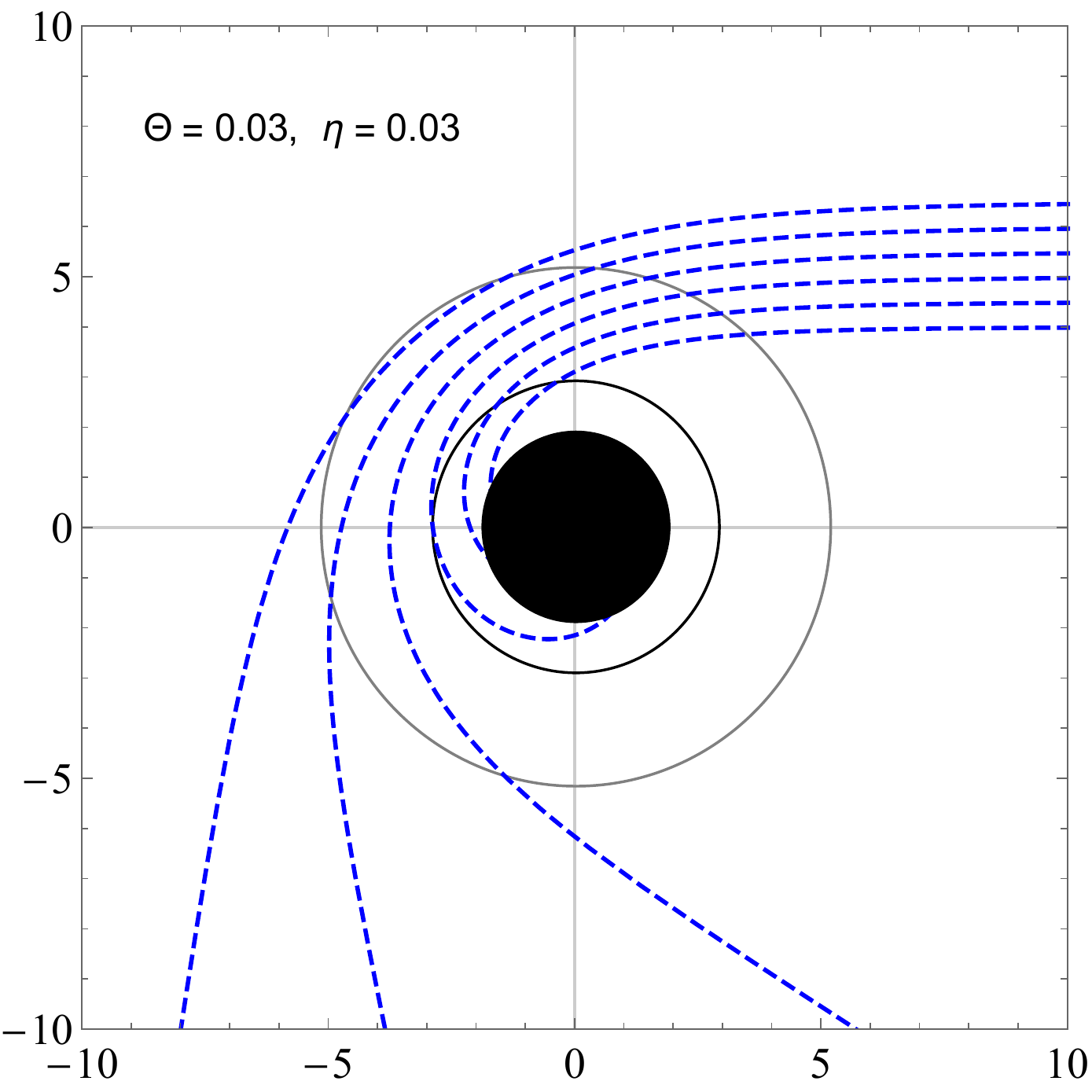}\label{geotha}}
 \qquad
 \subfigure[]{\includegraphics[scale=0.35]{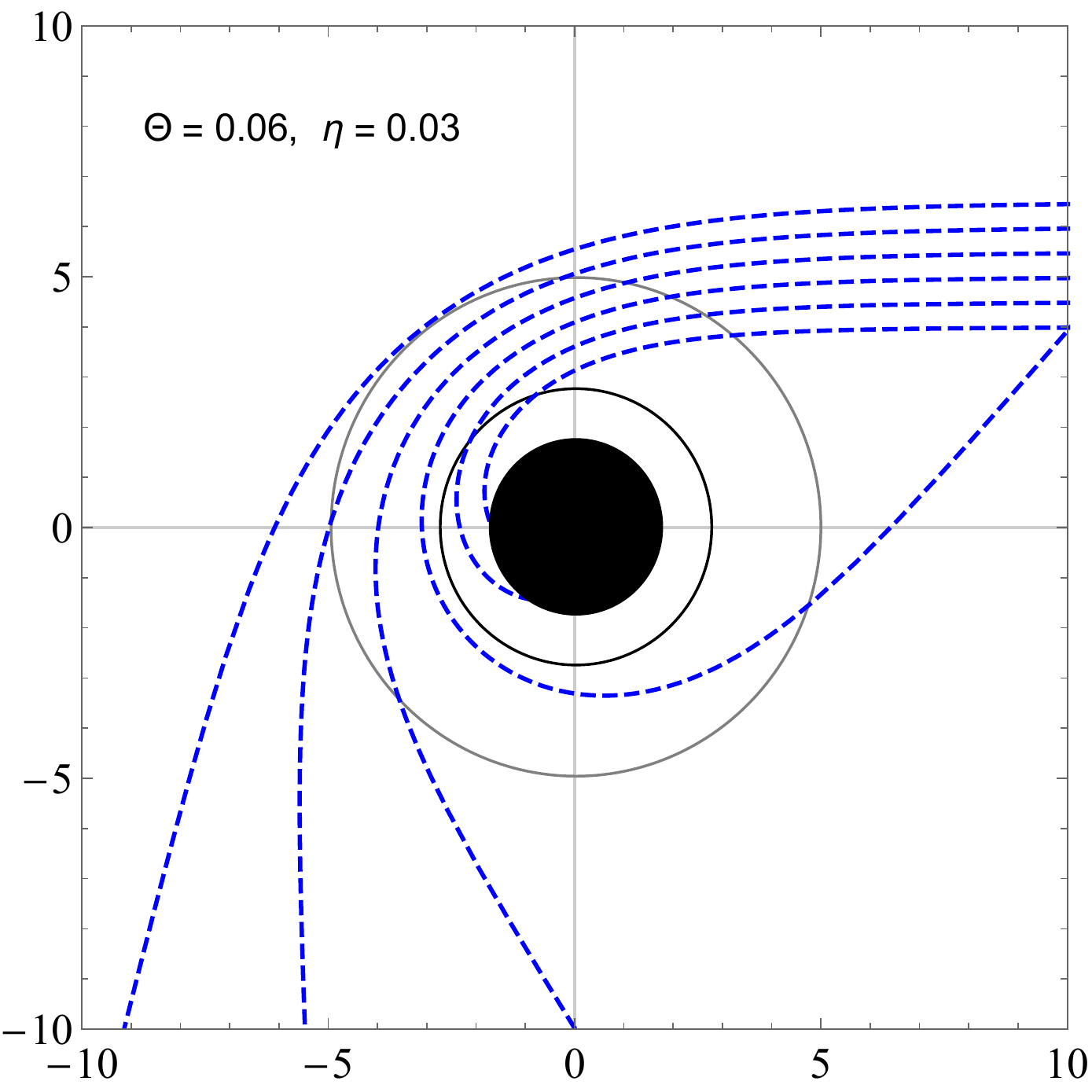}\label{geothb}}
 \qquad
 \subfigure[]{\includegraphics[scale=0.35]{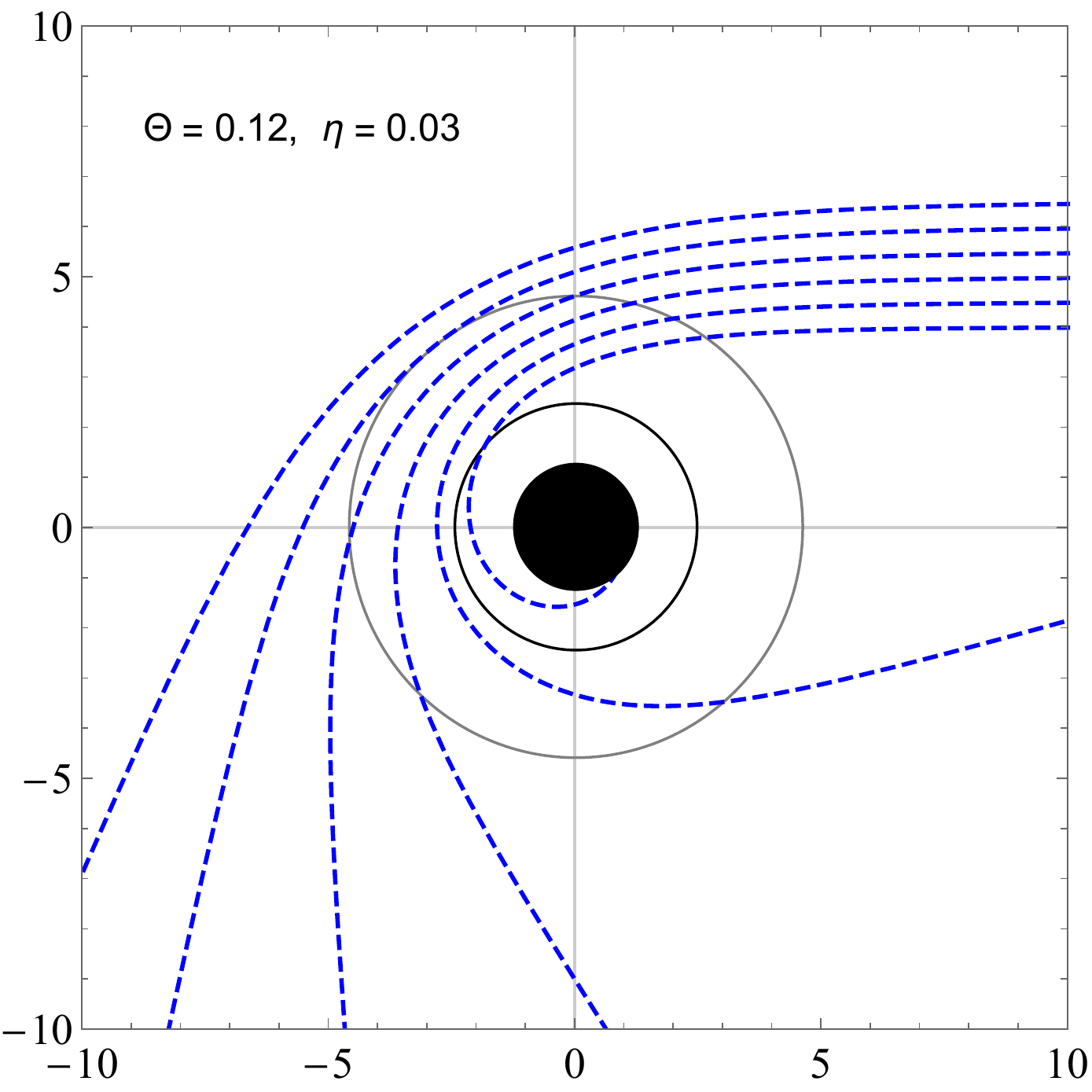}\label{geothc}}
 \qquad
 \subfigure[]{\includegraphics[scale=0.35]{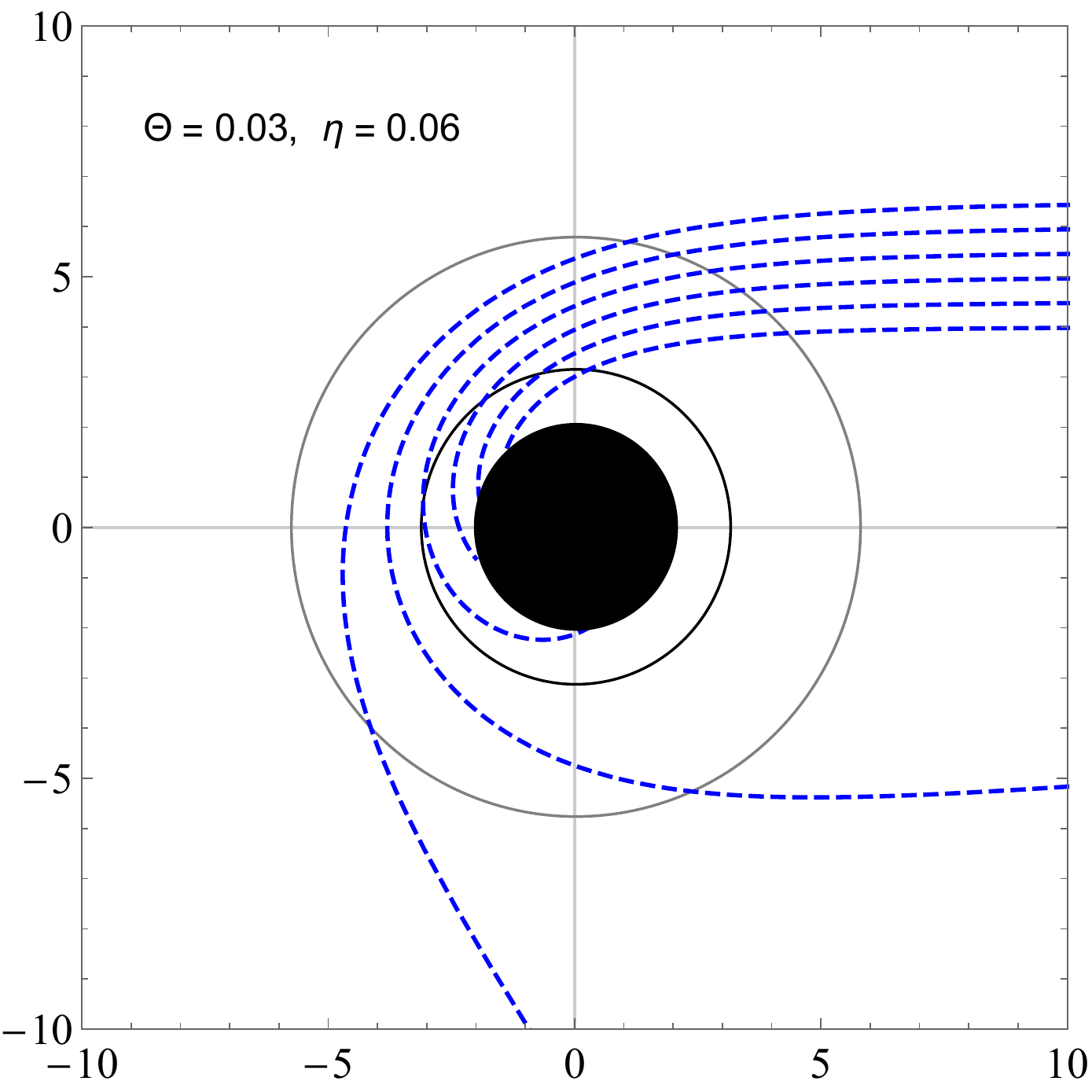}\label{geothd}}
 \qquad
 \subfigure[]{\includegraphics[scale=0.35]{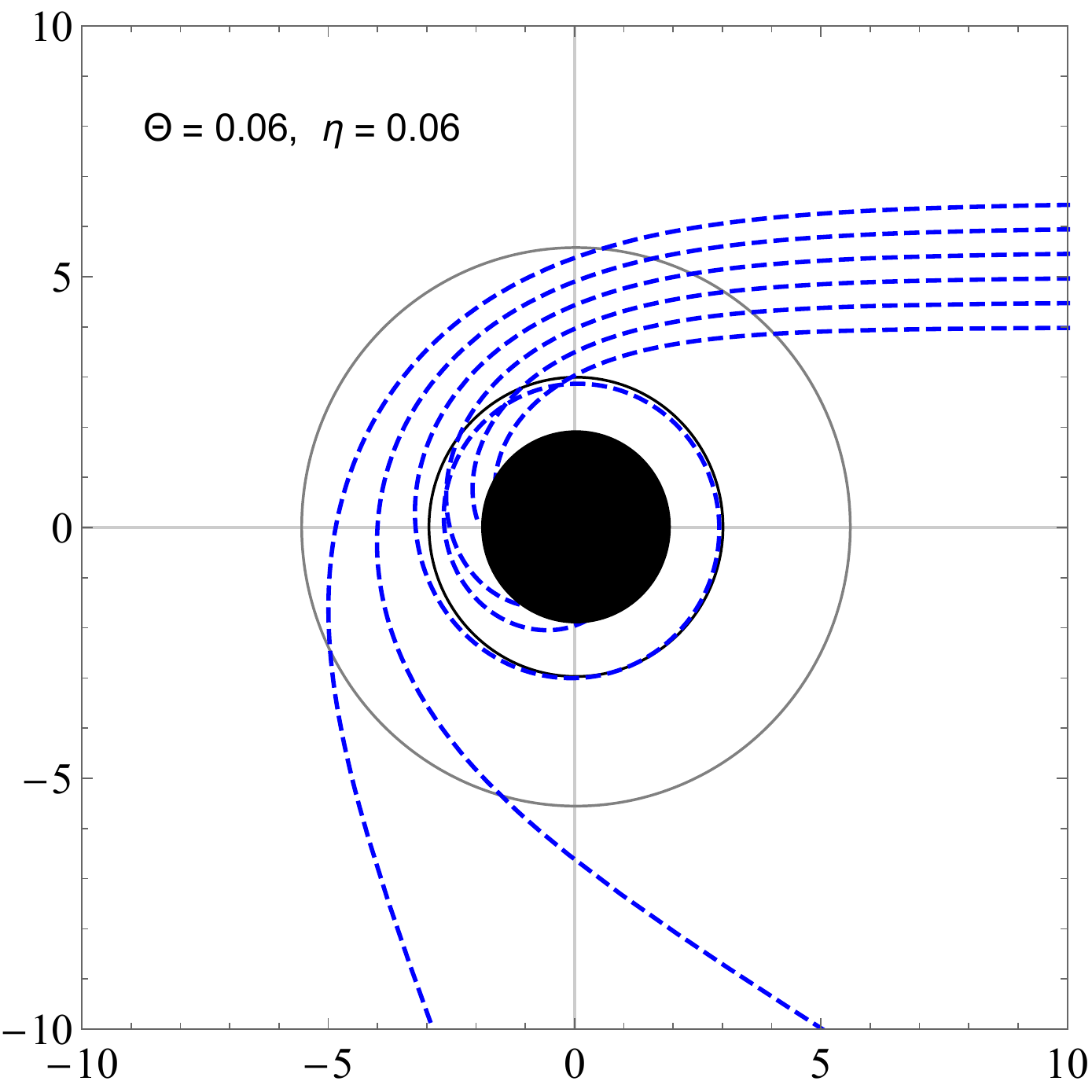}\label{geothe}}
 \qquad
 \subfigure[]{\includegraphics[scale=0.35]{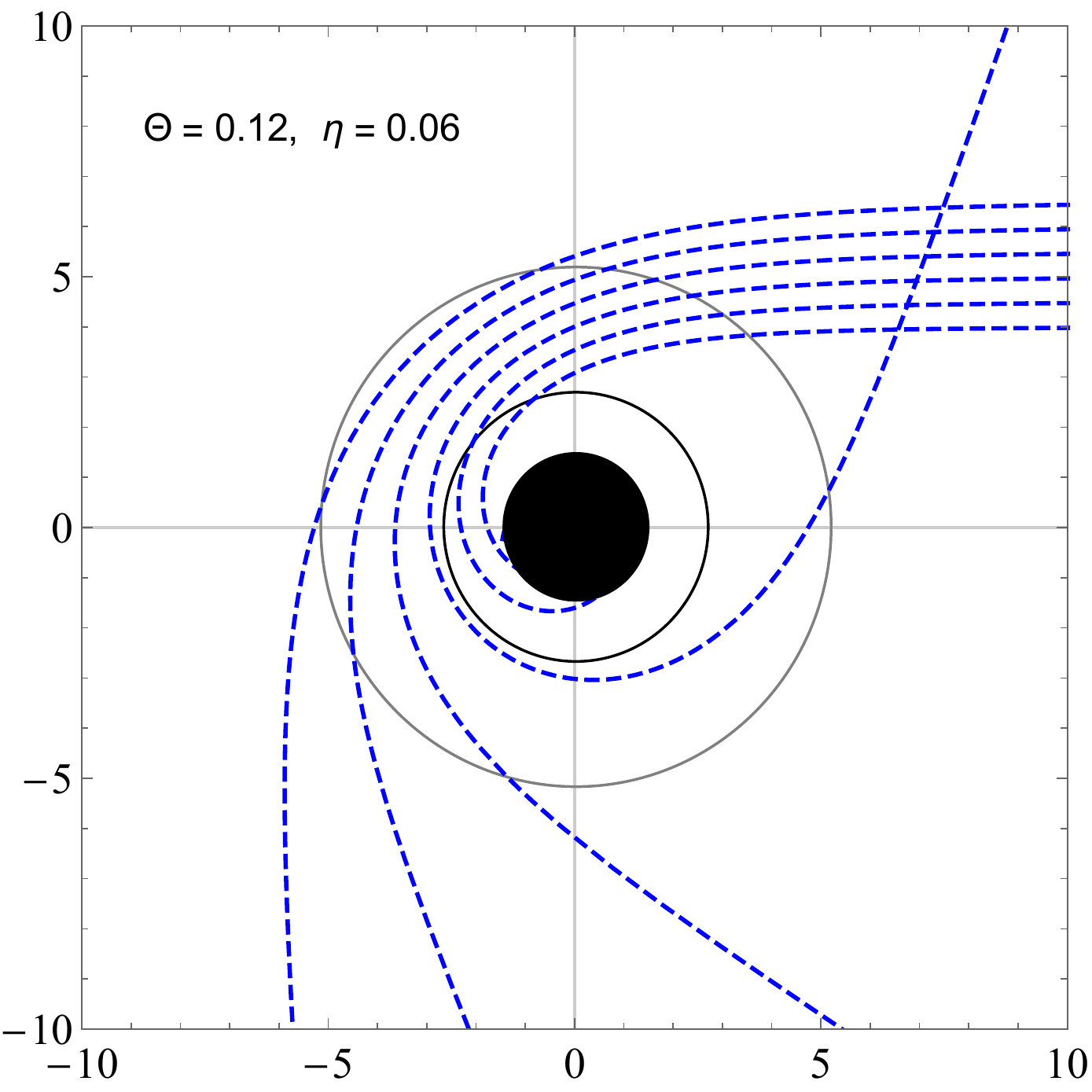}\label{geothf}}
 \caption{Null geodesics in polar coordinates for different values of $ \Theta $ and $ \eta $, assuming values for the impact parameter between 4 and 7. The radius of the horizon corresponds to the radius of the disk in black, the inner black ring is the critical radius or the radius of the sphere of photos while the outer ring corresponds to the critical impact parameter.} 
 \label{geot}
\end{figure}

As a starting point we consider the following Lagrangian density:
\begin{eqnarray}
\mathcal{L} \equiv \dfrac{1}{2}g_{\mu\nu}\dot{x}^{\mu}\dot{x}^{\nu},
\end{eqnarray}
being ``." the derivative with respect to an affine parameter.
Hence, by applying the line element (\ref{metrncm}), we have
\begin{equation}
2\mathcal{L} = F(r)\dot{t}^{2} - \dfrac{\dot{r}^{2}}{F(r)} - r^{2}\left( \dot{\vartheta}^{2} + \sin^{2}\vartheta \dot{\phi}^{2} \right).
\label{elidot}
\end{equation} 
At this point, we will consider the motion on the equatorial plane by taking $ \vartheta=\pi/2$. 
Hence, we can find the equations of motion from the Hamilton-Jacobi equation such that
\begin{equation}
E = F(r)\dot{t}, \qquad \quad L = r^{2}\dot{\phi},
\label{EL}
\end{equation}
where $ E $ (energy) and $ L $ (angular momentum) are the conserved quantities.
In the null geodesics analysis we have, $g_{\mu\nu}\dot{x}^{\mu}\dot{x}^{\nu} = 0$ and so we find
\begin{equation}
\dot{r}^{2}  + F(r)\dfrac{L^{2}}{r^{2}} = E^{2}.
\label{eqEner}
\end{equation}
Now, by introducing a new variable $ u=1/r $ to write the orbital equation 
in terms of the radii $ r_{\eta} $ and $ r_{\theta} $, we obtain
\begin{equation}
\left(\dfrac{du}{d\phi}\right)^2 =\dfrac{1}{b^2} - \left(1-8\pi\eta^2\right) \left(1 - u r_{\eta}\right) \left(1 - u r_{\theta}\right)u^{2},
\label{eqD1}
\end{equation}
here $b = L/E$  is the impact parameter defined as the perpendicular distance (measured at infinity) between the geodesic and a parallel line that passes through the origin. 
Hence, by differentiating Eq. \eqref{eqD1}, we find
\begin{equation}
\dfrac{d^{2}u}{d\phi^{2}} = \dfrac{\left(1-8\pi\eta^2\right)u}{2}\left[u r_{\eta}\left(1 - u r_{\theta}\right) + u r_{\theta}\left(1 - u r_{\eta}\right) - 2\left(1 - u r_{\eta}\right) \left(1 - u r_{\theta}\right)\right].
\label{eqD2}
\end{equation}
The behavior of the geodesic lines for different values of the impact parameter $ b $ can be obtained by solving the equations 
\eqref{eqD1} and \eqref{eqD2} numerically.
In Fig.~\ref{geot} we present the change of the geodesic lines for different values of the parameters $b$, $\Theta$ and $\eta$.
In the graphics, we have a black disk that represents the limit of the event horizon, the internal dotted circle is the radius for the photon sphere (critical radius), and the external dashed circle is the critical impact parameter (shadow). 
However, by fixing $\eta$ and increasing the values of the parameter $\Theta$ we see a reduction in the effect of the black hole on the light beam, while for fixed $\Theta$ and varying $\eta$ we have an increase in this effect.


To determine the critical radius ($ r_c $) and the critical impact parameter ($ b_c $), we impose conditions 
$du/d\phi = 0$ and $d^2u/d\phi^2 = 0$, to obtain
\begin{eqnarray}
r_{c} &=& \dfrac{1}{4}\left[ 3\left(r_{\eta} + r_{\theta}\right) 
+ \sqrt{9\left(r_{\eta} + r_{\theta}\right)^2 - 32r_{\eta}r_{\theta}}\right], \\
b_{c} &=& \dfrac{r_{c}^2}{\sqrt{\left(1-8\pi\eta^2\right)\left(r_{c} - r_{\eta}\right) \left(r_{c} - r_{\theta}\right)}}.
\label{paramcrit}
\end{eqnarray} 
Therefore, for high frequencies the absorption is determined by
\begin{equation}
\sigma_{abs}^{hf}=\pi b^2_c = \dfrac{\pi\left( 3M + \sqrt{9M^2 + K_1} \right)^4}
{8\left(1-8\pi\eta^2\right)^3\left( 3M^2 + K_2 + M\sqrt{9M^2 + K_1} \right)},
\end{equation}
where 
\begin{eqnarray}
&&K_1=-\frac{64M\left(1-8\pi\eta^2\right)\sqrt{\theta}}{\sqrt{\pi}} 
+ \frac{128\left(1-8\pi\eta^2\right)^2{\theta}}{{\pi}},
\\
&&K_2=-\dfrac{16M\left(1-8\pi\eta^2\right)\sqrt{\theta}}{\sqrt{\pi}}
+\dfrac{32\left(1-8\pi\eta^2\right)^2 \theta}{\pi}.
\end{eqnarray}
Observe that making $ \eta=\theta=0 $, we have $ \sigma_{abs}^{hf}=27\pi M^2  $ which is the result for the Schwarzschild black hole. 

For $ \eta\neq 0 $ and $ \theta=0 $, the absorption becomes
\begin{eqnarray}
\sigma_{abs}^{hf}=27\pi M^2(1+24\pi\eta^2 + \cdots).
\end{eqnarray}
Hence by varying the values of $ \eta $, the absorption value at the high frequency limit has its value increased as shown in Fig.~\ref{absta}.
For $ \eta=0 $ and $ \theta\neq 0 $, we obtain
\begin{eqnarray}
\sigma_{abs}^{hf}=27\pi M^2 - 72\pi M\sqrt{\frac{\theta}{\pi}}+\cdots.
\end{eqnarray}
Here the absorption at the high frequency limit has its value reduced when we vary $ \theta $ as shown in 
Fig.~\ref{abstb} .

Now, we will calculate the size of the black hole shadow using the geodetic study and expressing it via celestial coordinates.
\begin{eqnarray}
\alpha &=& \lim\limits_{r_{o} \to \infty}\left[- r_{o}^{2} \sin\theta_{o}\dfrac{d\phi}{dr}\Bigr\rvert_{\theta = \theta_{o}} \right] ,\label{coodcelesta}\\
\beta &=& \lim\limits_{r_{o} \to \infty}\left[r_{o}^{2} \dfrac{d\theta}{dr}\Bigr\rvert_{\theta = \theta_{o}}\right] ,
\label{coodcelestb}
\end{eqnarray}
where $\left(r_{o}, \theta_{o}\right)$ is the observer's position at infinity. 
In this case, for an observer in the equatorial plane, that is in $ \theta_o = \pi/2 $, we have the following relation
\begin{equation}
R_{s} \equiv \sqrt{\alpha^{2} + \beta^{2}} = b_{c}. 
\label{rshw}
\end{equation}
Then
\begin{eqnarray}
R_{s}=\left[ \dfrac{\left( 3M + \sqrt{9M^2 + K_1} \right)^4}
{8\left(1-8\pi\eta^2\right)^3\left( 3M^2 + K_2 + M\sqrt{9M^2 + K_1} \right)} \right]^{1/2}.
\label{rs}
\end{eqnarray}
\begin{figure}[!htb]
 \centering
 \subfigure[]{\includegraphics[scale=0.5]{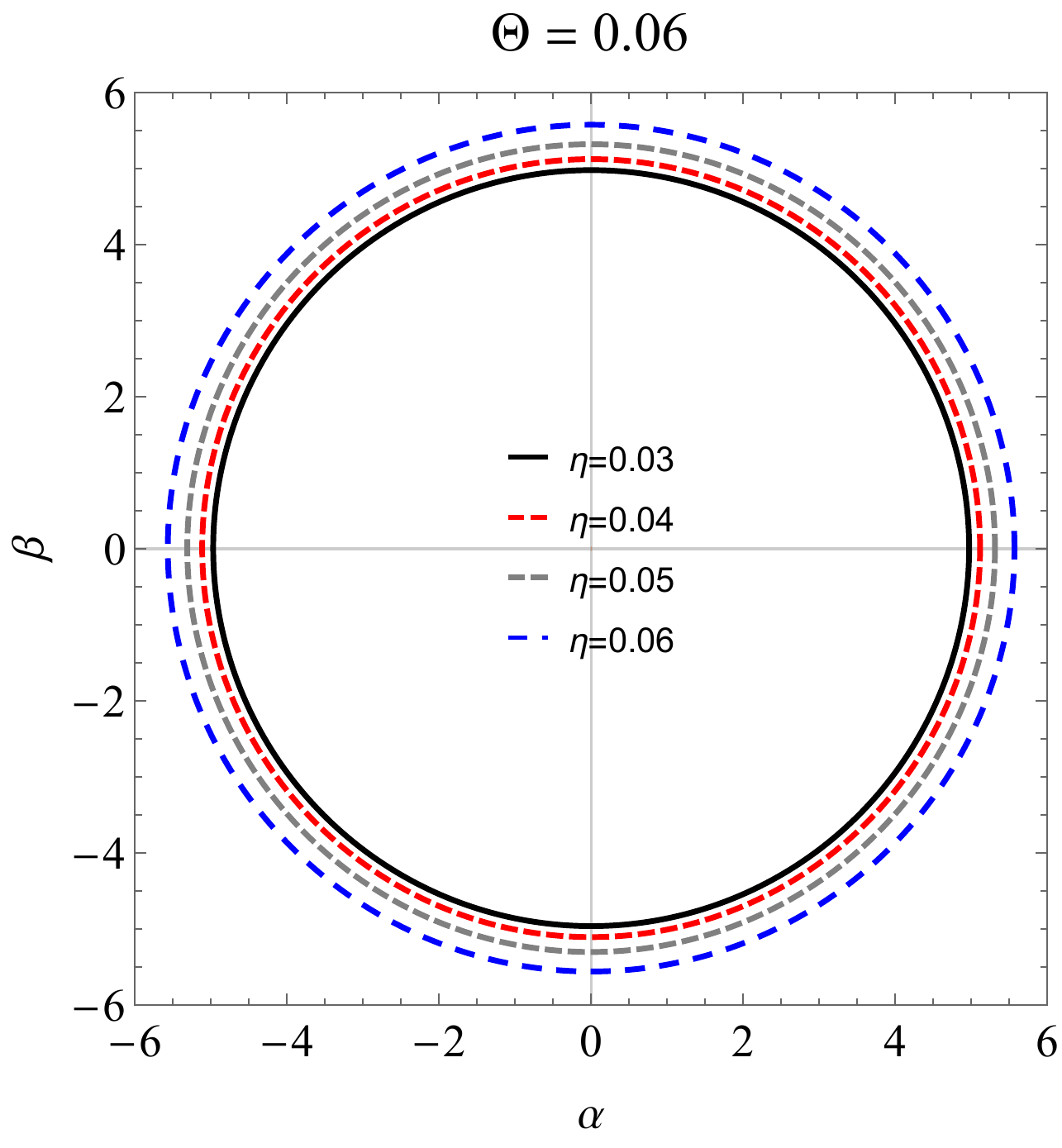}\label{shwa}}
 \qquad
 \subfigure[]{\includegraphics[scale=0.5]{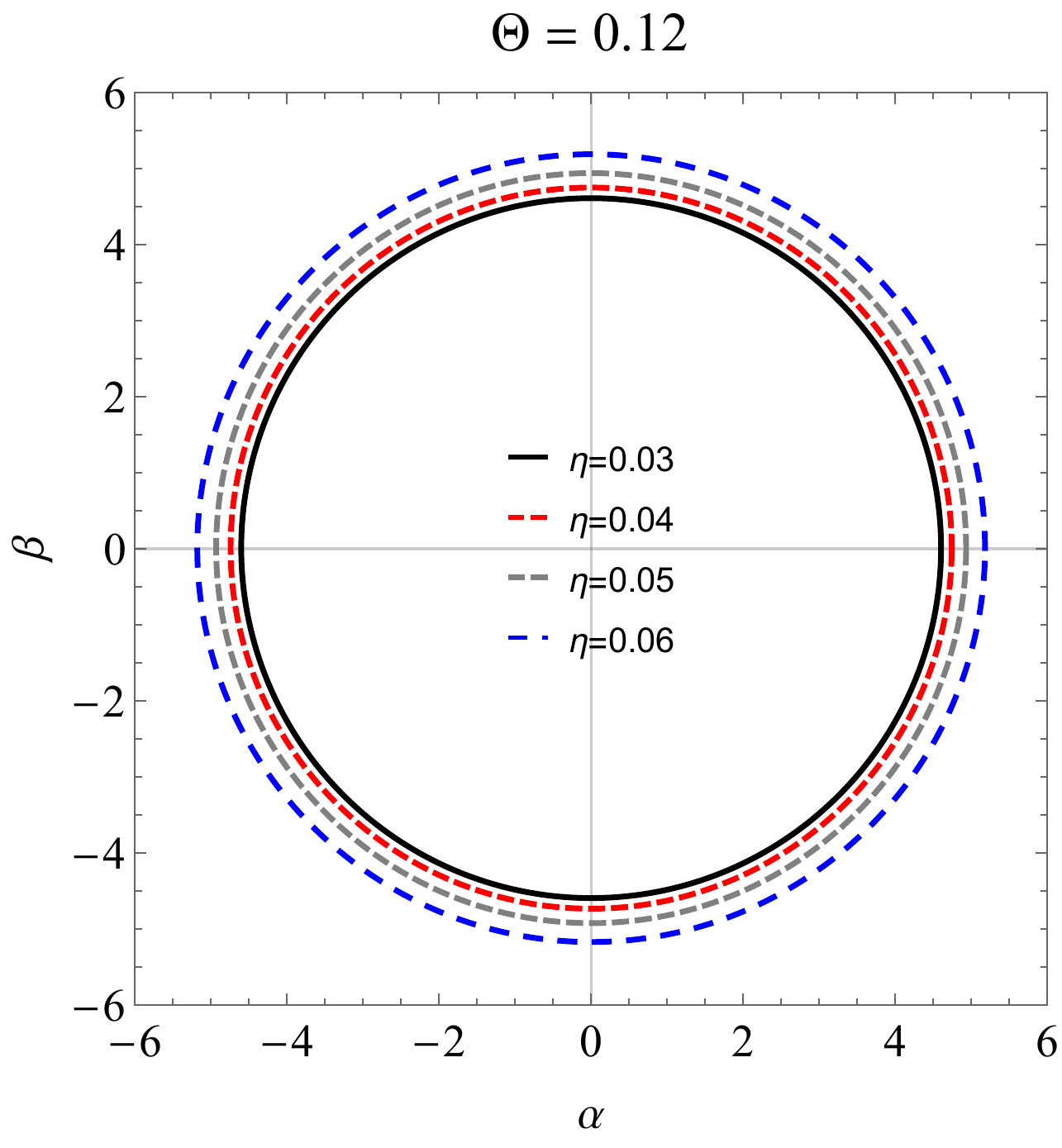}\label{shwb}}
 \caption{Influence of the $ \Theta $ and $  \eta$ parameters on the shadow radius.} 
 \label{shwt}
\end{figure}
For $\theta$ and $\eta$ small, we get
\begin{eqnarray}
R_s=3\sqrt{3}M\left(1+12\pi\eta^2\right) - 4\sqrt{3}(1+4\pi\eta^2)\sqrt{\frac{\theta}{\pi}}+\cdots .
\label{rsp}
\end{eqnarray}
Note that the global monopole causes an increase in the shadow radius when we increase the $\eta$ parameter as we can see in 
Fig.~\ref{shwt}. On the other hand, noncommutativity has the effect of reducing the shadow radius. 
Besides, for $ \theta=0 $ in (\ref{rsp}), we have
\begin{eqnarray}
R_s\approx 3\sqrt{3}M\left(1+12\pi\eta^2\right).
\label{rsp1}
\end{eqnarray}
Here the shadow radius has its value increased when we vary the $\eta $ parameter.

For $ \eta=0 $ in (\ref{rsp}), we obtain
\begin{eqnarray}
R_s=3\sqrt{3}M - 4\sqrt{3}\sqrt{\frac{\theta}{\pi}}+\cdots .
\end{eqnarray}
Hence the shadow radius has its value reduced when we vary the $\Theta $ parameter.

\begin{figure}[!htb]
 \centering
 \subfigure[]{\includegraphics[scale=0.3]{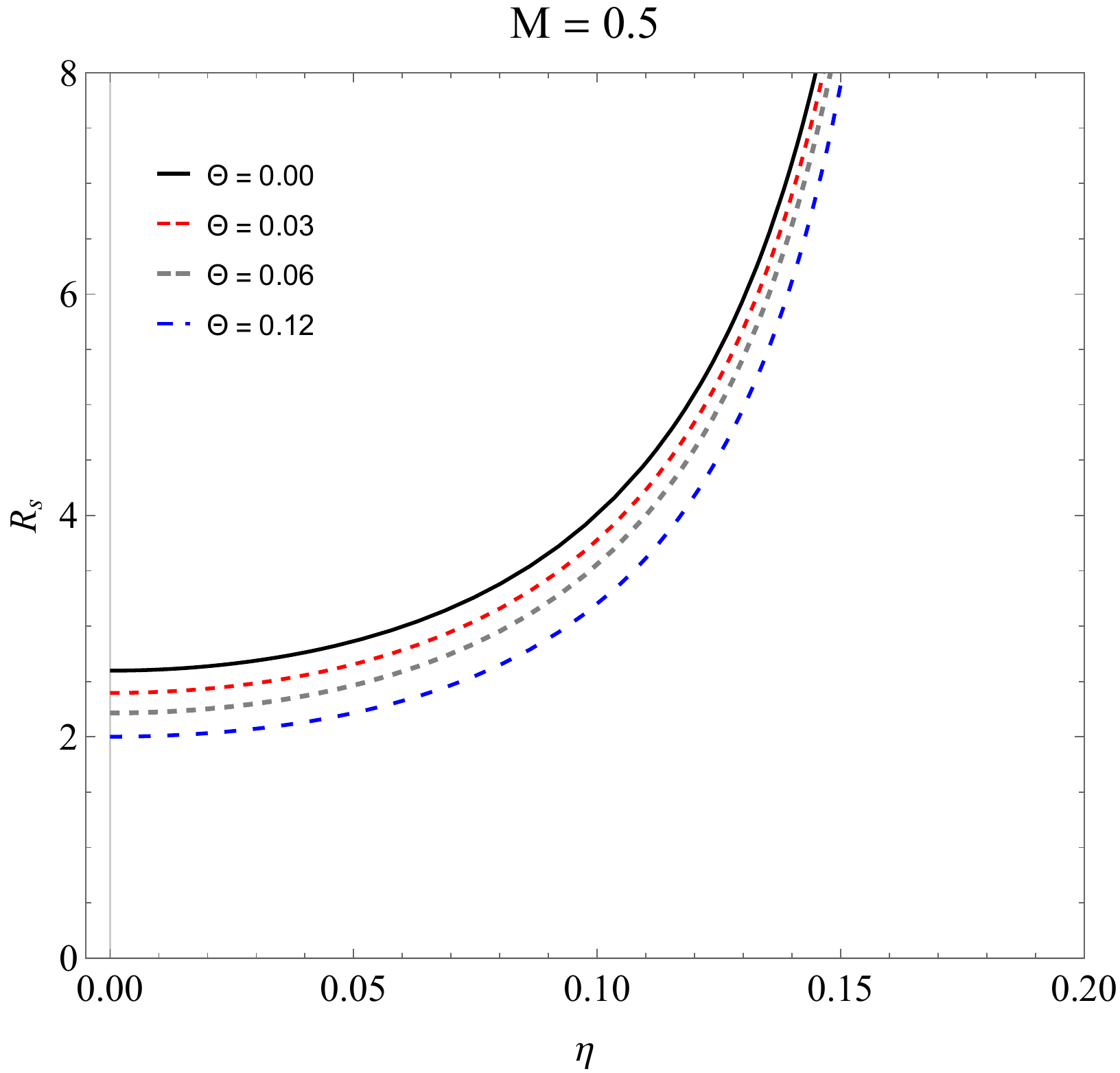}\label{rsm05}}
 \qquad
 \subfigure[]{\includegraphics[scale=0.3]{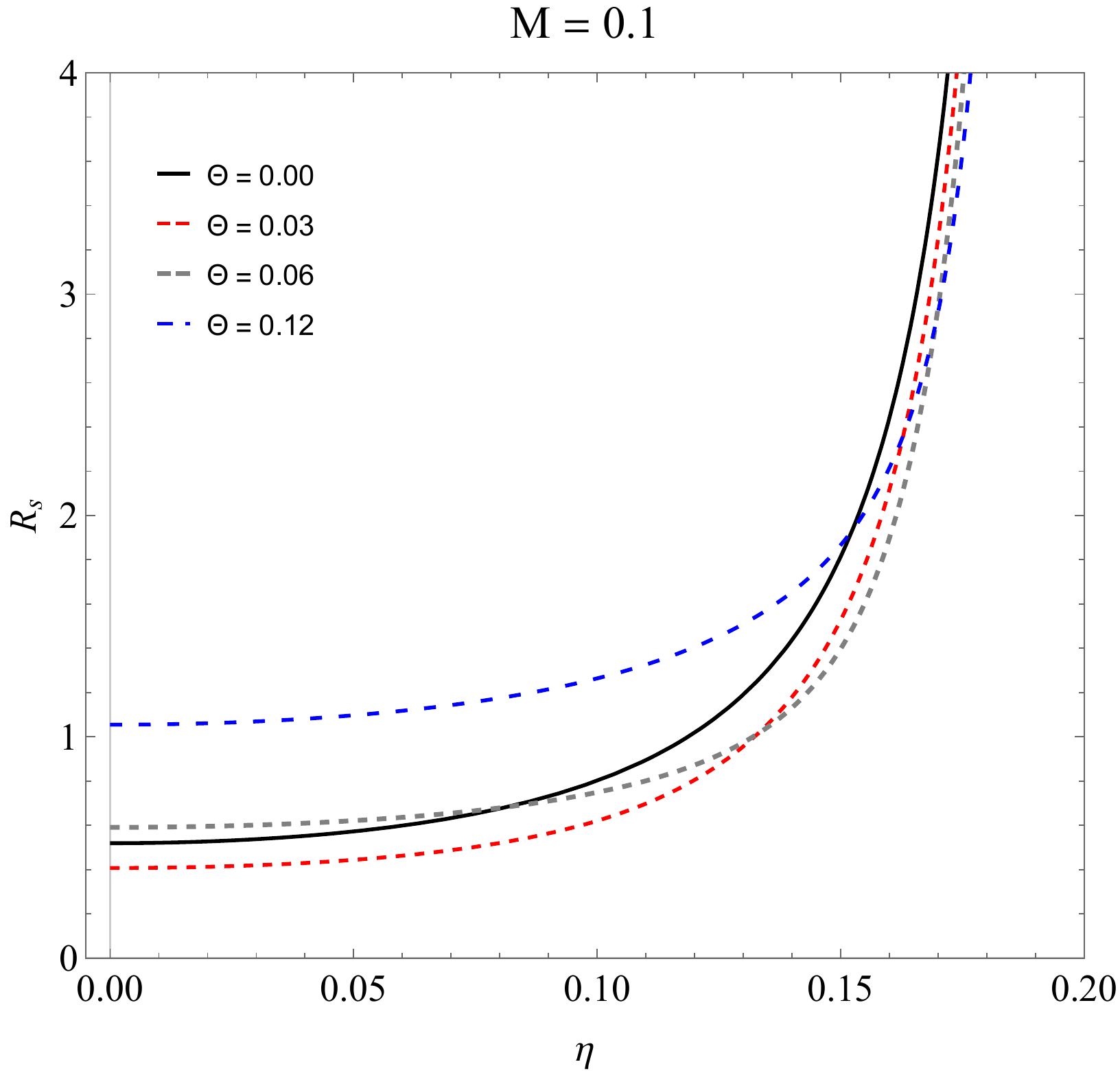}\label{rsm01}}
  \qquad
 \subfigure[]{\includegraphics[scale=0.3]{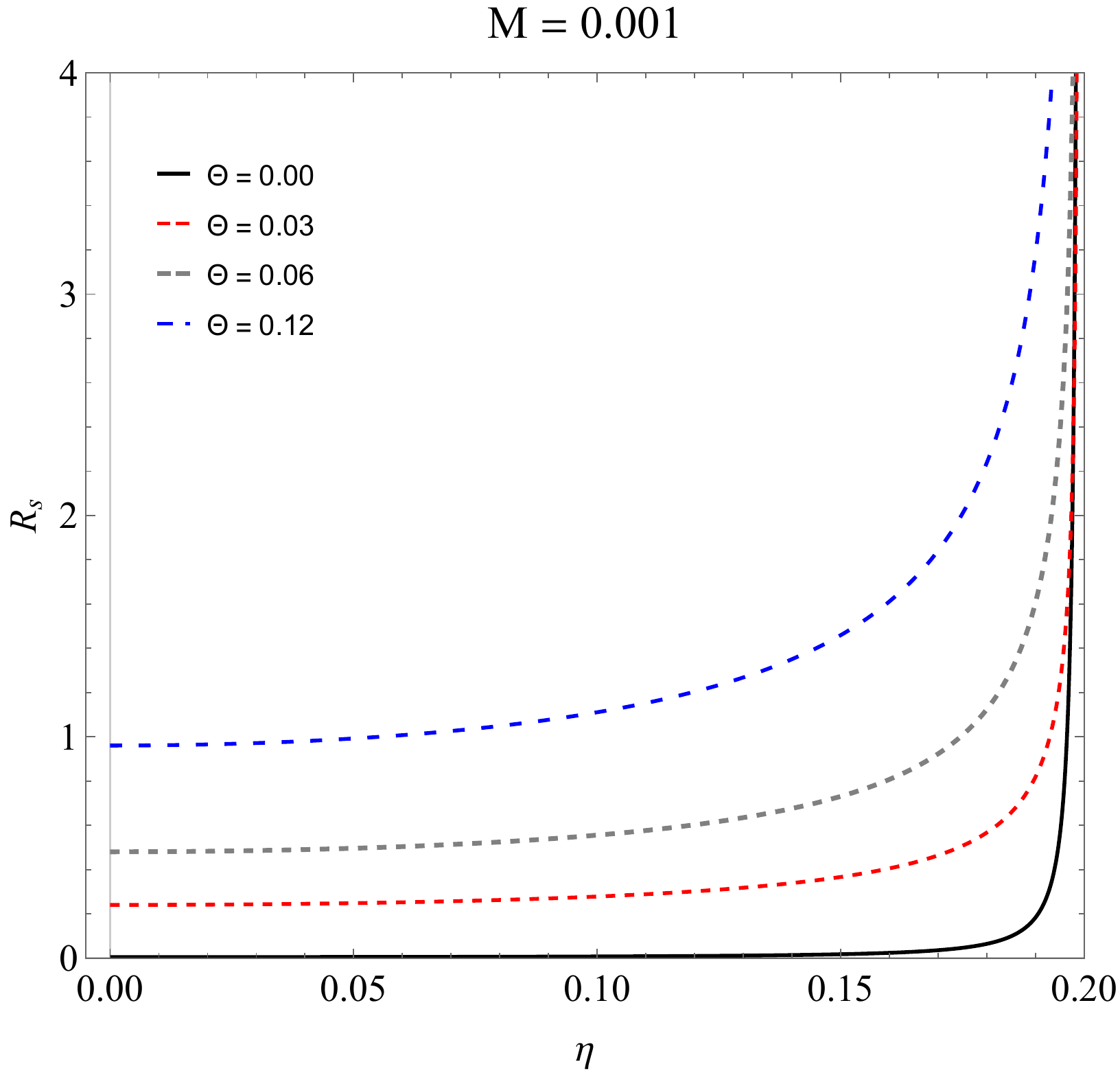}
 }
  \caption{\footnotesize{Influence of the $ \Theta $ and $  \eta$ parameters on the shadow radius.}} 
  \label{rsm}
\end{figure}

\begin{figure}[!htb]
 \centering
 \subfigure[]{\includegraphics[scale=0.36]{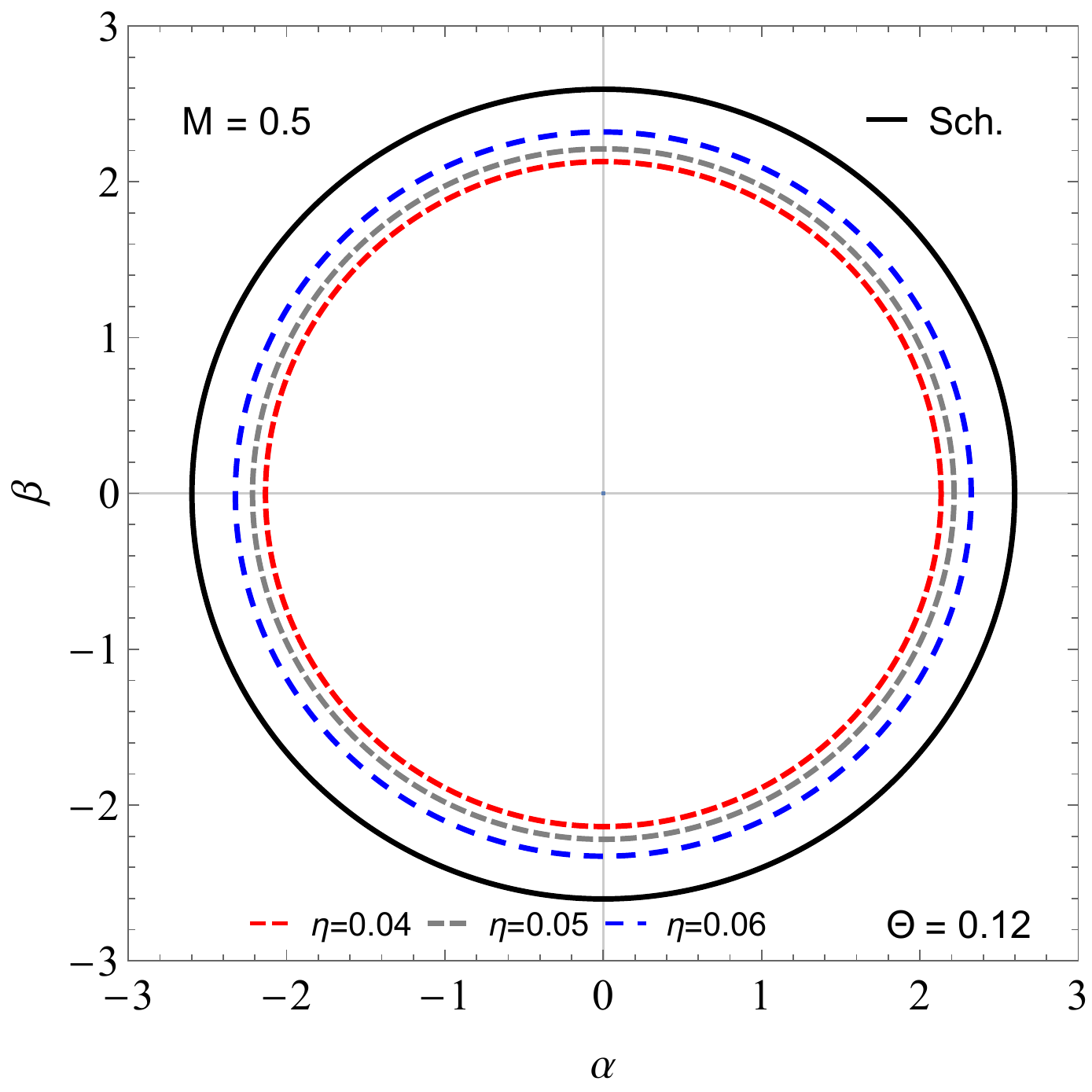}\label{sd05}}
 \qquad
 \subfigure[]{\includegraphics[scale=0.36]{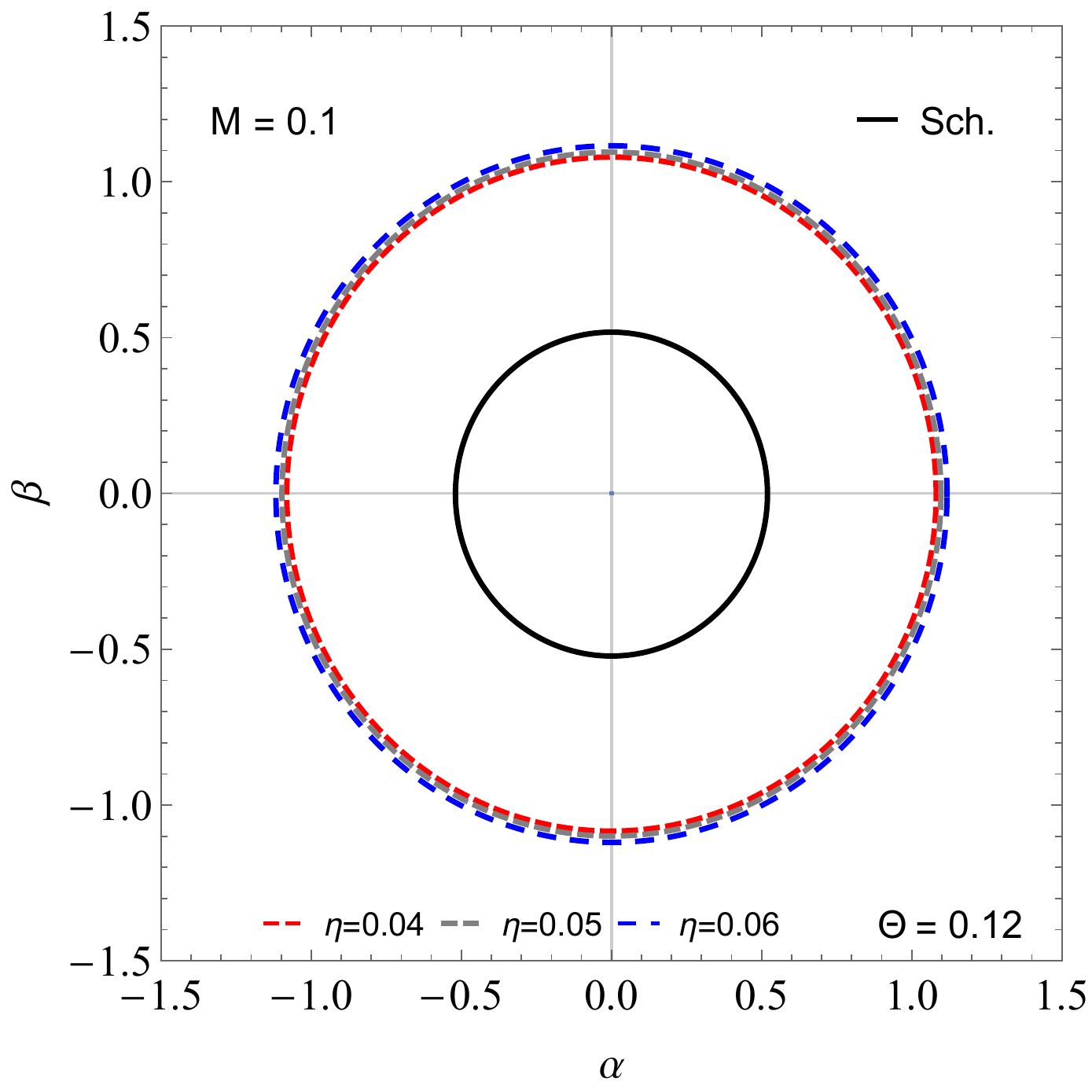}\label{sd01}}
  \qquad
 \subfigure[]{\includegraphics[scale=0.36]{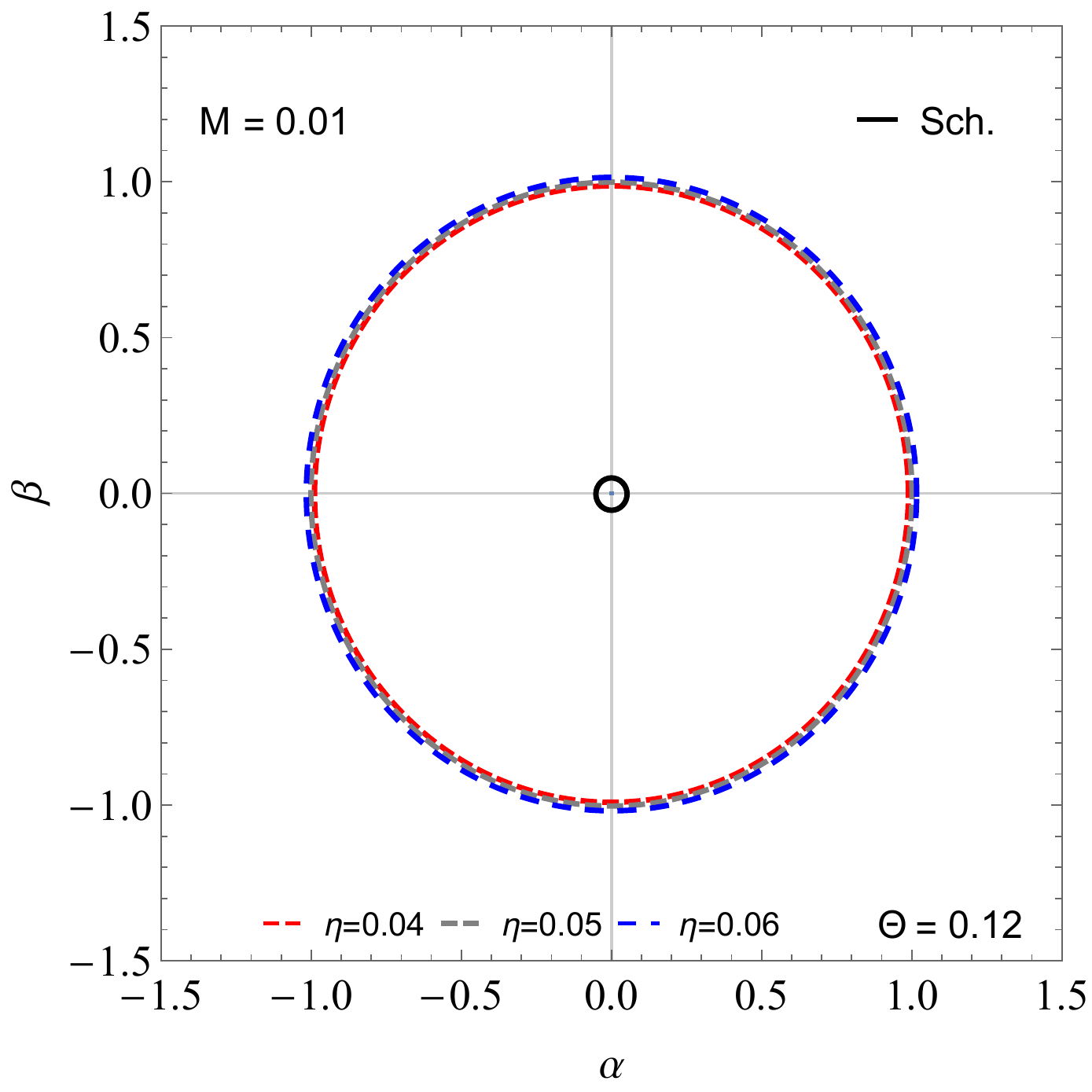}\label{sd001}}
  \caption{\footnotesize{Influence of the $ \Theta $ and $  \eta$ parameters on the shadow radius.}} 
  \label{sd}
\end{figure}

Next, taking the null mass limit in (\ref{rs}) we find that the shadow radius is non-zero, given by
\begin{eqnarray}
R_{smin}\approx \frac{4}{3}(1+12\pi\eta^2)M_{min}.
\end{eqnarray}
Therefore, we can see that by varying the $\eta$ parameter with a fixed $\Theta$ the shadow radius increases as we can see in 
Fig.~\ref{rsm} and Fig.~\ref{sd}.

\subsection{Classic Analysis}
Rewriting the equation of the orbit \eqref{eqD1} in terms of $r$, we have
\begin{equation}
\dfrac{d\phi}{d r} = \dfrac{1}{r^{2}}\left(\dfrac{1}{b^{2}} - \dfrac{(1-8\pi\eta^{2})}{r^{2}} + \dfrac{2M}{r^{3}}-\dfrac{8M}{r^{4}}\sqrt{\dfrac{\theta}{\pi}}\right)^{-1/2}.
\label{eq_geo}
\end{equation}
To calculate the total angle deflected by the beam we have to integrate the above equation. One way to integrate analytically is to take the limit $M/r << 1$.  Before doing this expansion let us organize the equation by making  changes of variables
\begin{eqnarray}
y &=& \dfrac{1}{r}\left(\delta - \dfrac{M}{r} + \dfrac{(\tilde{\theta} - 1)}{2}\dfrac{M^{2}}{r^{2}}\right), \\
y^{2} &=& \dfrac{1}{r^{2}}\left[ \left(\delta - \dfrac{M}{r}\right)^{2} + \left(\delta - \dfrac{M}{r}\right)\dfrac{(\tilde{\theta} - 1)M^{2}}{r^{2}} + \dfrac{(\tilde{\theta} - 1)^{2}}{4}\dfrac{M^{4}}{r^{4}}\right], \\
\dfrac{dy}{dr} &=& -\dfrac{\delta}{r^{2}} + \dfrac{2M}{r^{3}} - \dfrac{3(\tilde{\theta} - 1)M^{2}}{2 r^{4}},
\end{eqnarray}
where we define $\tilde{\theta} = \dfrac{8}{M}\sqrt{\dfrac{\theta}{\pi}}$ and $\delta = 1-8\pi\eta^{2} $, so the equation \eqref{eq_geo} is rewritten in this new variable as
\begin{eqnarray}
\small
\dfrac{d\phi}{dy} &=& \left(-\delta + \dfrac{2M}{r} - \dfrac{3(\tilde{\theta} - 1)}{2}\dfrac{M^{2}}{r^{2}}\right)^{-1}\left\lbrace \dfrac{1}{b^{2}} - \dfrac{y^{2}}{\delta^{2}}  + \dfrac{1}{\delta r^{2}}\left[(\delta -1)\left(-\delta + \dfrac{2M}{r} - \dfrac{\tilde{\theta}M^{2}}{r^{2}}\right)\right.\right.
\nonumber\\
&-&\left.\left.\dfrac{M^{2}}{\delta r^{2}}\left( \delta - 1 + (\tilde{\theta}-1)\dfrac{M}{\delta r} - (\tilde{\theta}-1)^{2}\dfrac{M^{2}}{4 \delta r^{2}}\right)\right]\right\rbrace^{-1/2}.
\end{eqnarray}

Now expanding the above equation at the limit $M/r << 1$ and organizing the terms, we find
\begin{small}
\begin{eqnarray}
\dfrac{d\phi}{dy} = \dfrac{b}{\delta^{2}\sqrt{1- \dfrac{y^{2}b^{2}}{\delta^{2}}}}\left\lbrace \delta + \dfrac{2M}{r} - \dfrac{2\left(-4 + 3\delta(\tilde{\theta}-1) \right)}{\delta^{2}}\dfrac{M^{3}}{r^{3}} +\dfrac{\left(8 - 3(\tilde{\theta} - 1)\delta \right)}{2\delta}\dfrac{M^{2}}{r^{2}} + \dfrac{b^{2}\delta^{3}(1-\delta)}{2\left(b^{2}y^{2} - \delta^{2} \right)r^{2}}  + \cdots\right\rbrace
.
\end{eqnarray}
\end{small}
We can identify the $y$ in terms of $M/r$ and a $y^{2}$ in $M^{2}/r^{2}$, so we have
\begin{equation}
\dfrac{d\phi}{dy} = \dfrac{b}{\sqrt{1- \dfrac{y^{2}b^{2}}{\delta^{2}}}} \left[ \delta + \dfrac{2My}{\delta}  + \dfrac{\left(12 - 3(\tilde{\theta}-1)\delta\right) M^{2}y^{2}}{2\delta^{3}}\right].
\end{equation}
The total angle deflected by the beam is given by 
\begin{small}
\begin{eqnarray}
\Delta\phi &=& 2 \int_{0}^{y_{0}} \dfrac{b}{\sqrt{1- \dfrac{y^{2}b^{2}}{\delta^{2}}}} \left[ \delta + \dfrac{2My}{\delta}  + \dfrac{\left(12 - 3(\tilde{\theta}-1)\delta\right) M^{2}y^{2}}{2\delta^{3}}\right]dy, \\
\Delta\phi &=&  - \dfrac{M\sqrt{1-b^{2}y_{0}^{2}/\delta^{2}}\left(8\delta^{2} - 3M y_{0} (4+\delta - \tilde{\theta})\right)}{2b\delta^{3}} +\dfrac{4b^{2}\delta^{2} + 3M^{2}(4 + \delta - \delta\tilde{\theta})\arcsin(by_{0}/\delta)}{2b^{2}\delta^{2}} + \dfrac{4M}{b \delta}.
\end{eqnarray}
\end{small}
Knowing that at the limit $M/r << 1$ we have that $y_{0}\rightarrow \delta/b$ and the result becomes
\begin{equation}
\Delta\phi = \dfrac{4M}{b\delta} + \pi + \dfrac{3\pi(4 + \delta(1 - \tilde{\theta}))M^{2}}{4b^{2}\delta^{2}}.
\end{equation}
Finally, the deflection angle $\chi = \Delta \phi - \pi$ in terms of $\sqrt{\theta}$ and $\eta^{2}$ is given by
\begin{equation}
\chi = \dfrac{4M}{b(1-8\pi\eta^{2})} + \dfrac{3\pi(5-8\pi\eta^{2}) M^{2}}{4b^{2}(1-8\pi\eta^{2})^{2}} - \dfrac{6\pi M}{b^{2}(1-8\pi\eta^{2})}\sqrt{\dfrac{\theta}{\pi}}.
\end{equation}

\begin{figure}[!htb]
 \centering
 \subfigure[]{\includegraphics[scale=0.35]{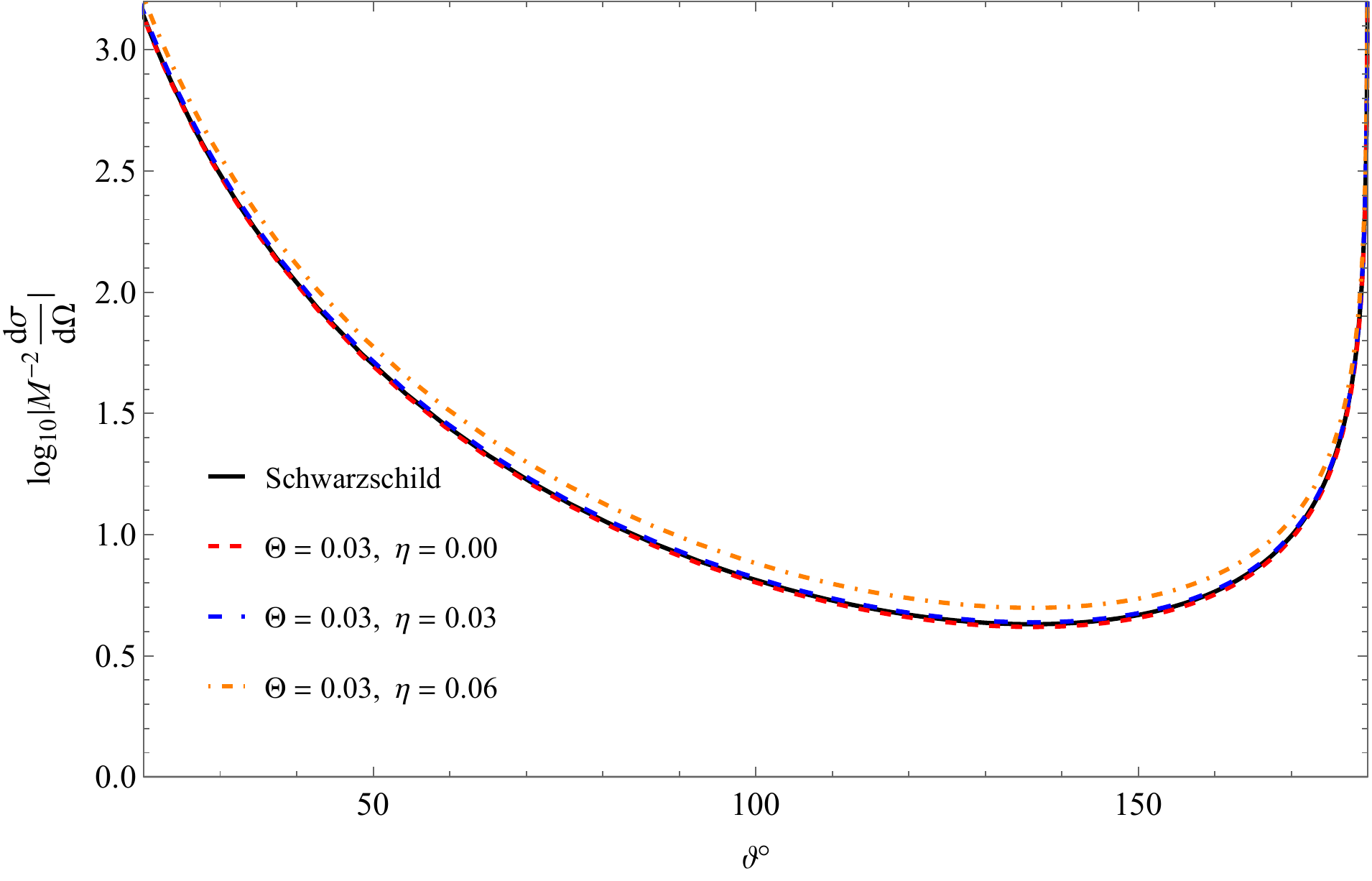}\label{Class}}
 \qquad
 \subfigure[]{\includegraphics[scale=0.38]{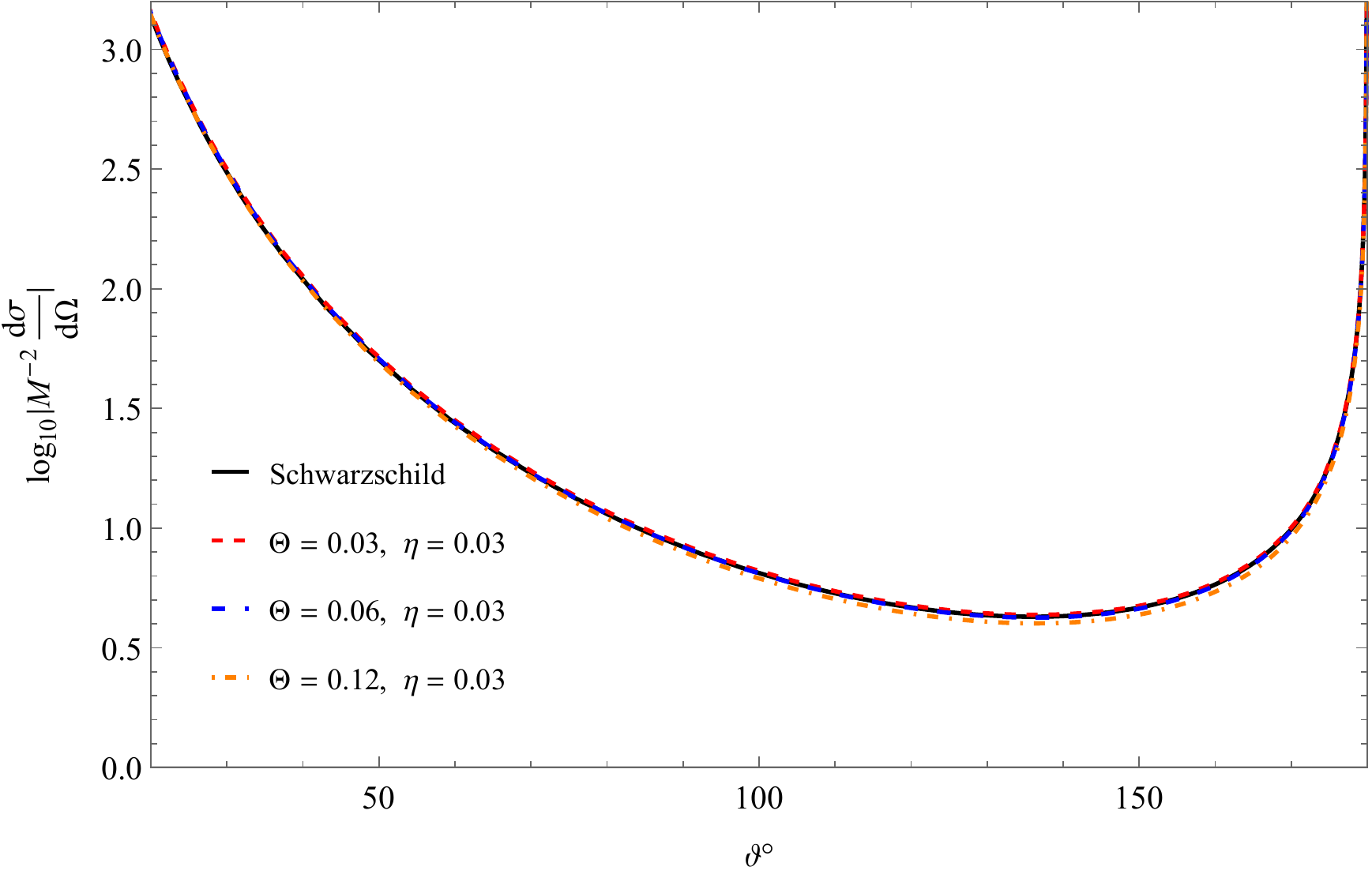}\label{ClassEta003}}
  \caption{\footnotesize{Classical scattering differential cross section.}} 
\end{figure}

With the relationship of the deflection angle and the impact parameter, we can combine this result with the classical differential scattering section equation
\begin{equation}
\dfrac{d \sigma_{clas}}{d \Omega} = \dfrac{b}{\sin\vartheta} \Big |\dfrac{db}{d\vartheta} \Big |,
\end{equation}
to get the following result at the limit of small $\vartheta$
\begin{eqnarray}
\dfrac{d \sigma_{clas}}{d \Omega} \approx \dfrac{16 M^{2}}{(-1 + 8\pi \eta^{2})^{2}\vartheta^{4}} + \dfrac{3\pi M^{2}(5-8\pi\eta^{2})}{4(-1 + 8\pi \eta^{2})^{2}\vartheta^{3}} - \dfrac{6\pi M}{(-1 + 8\pi \eta^{2})^{2}\vartheta^{3}}\sqrt{\dfrac{\theta}{\pi}}.
\label{scattclass}
\end{eqnarray}
We can see in the equation \eqref{scattclass} that the influence of the non-commutative parameter is very small Fig \ref{ClassEta003}, since the monopole term will have a contribution to the classical case even in the limit of small angles Fig \ref{Class} .

\begin{figure}[!htb]
 \centering
 \subfigure[]{\includegraphics[scale=0.35]{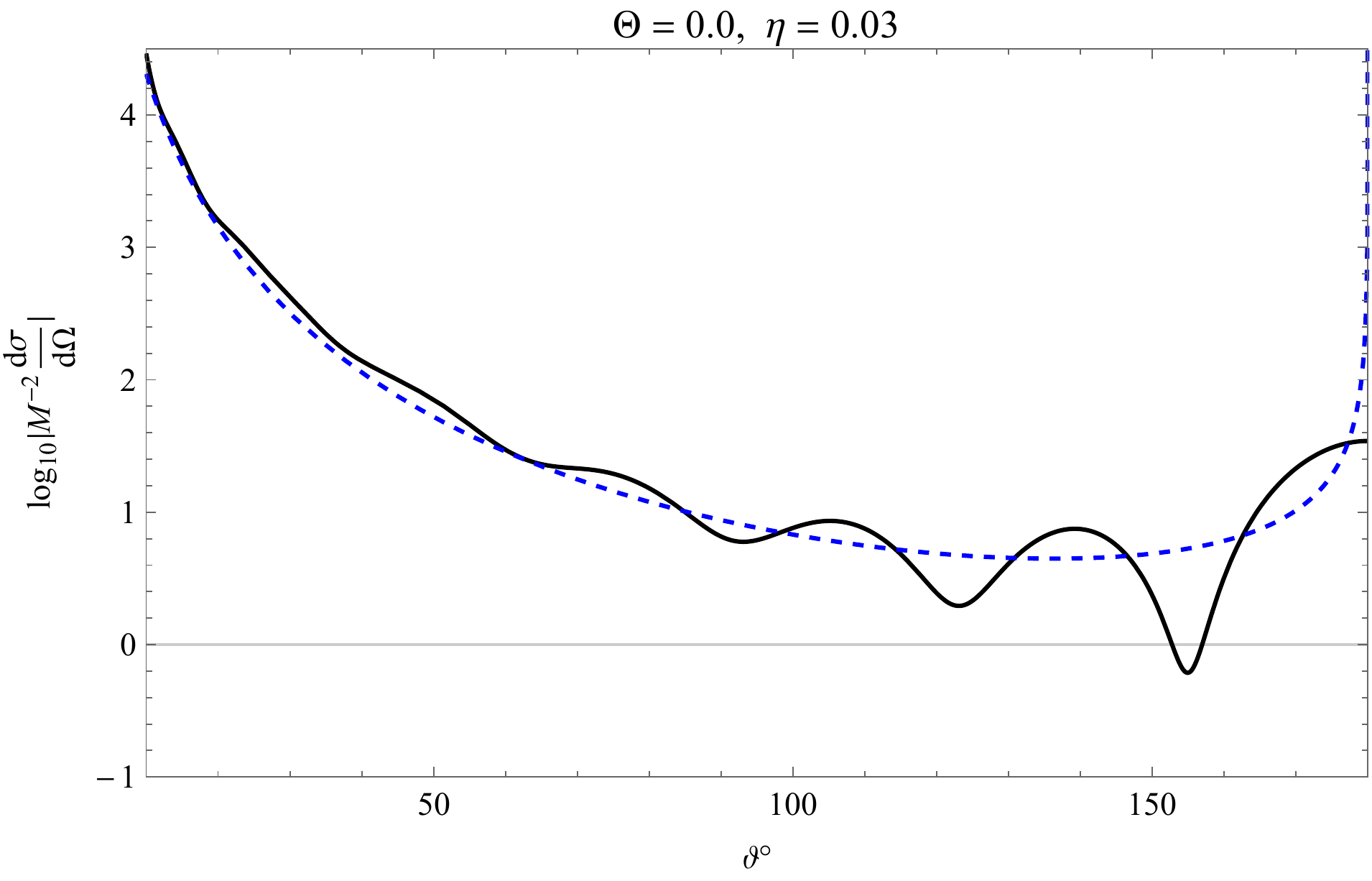}}
 \qquad
 \subfigure[]{\includegraphics[scale=0.35]{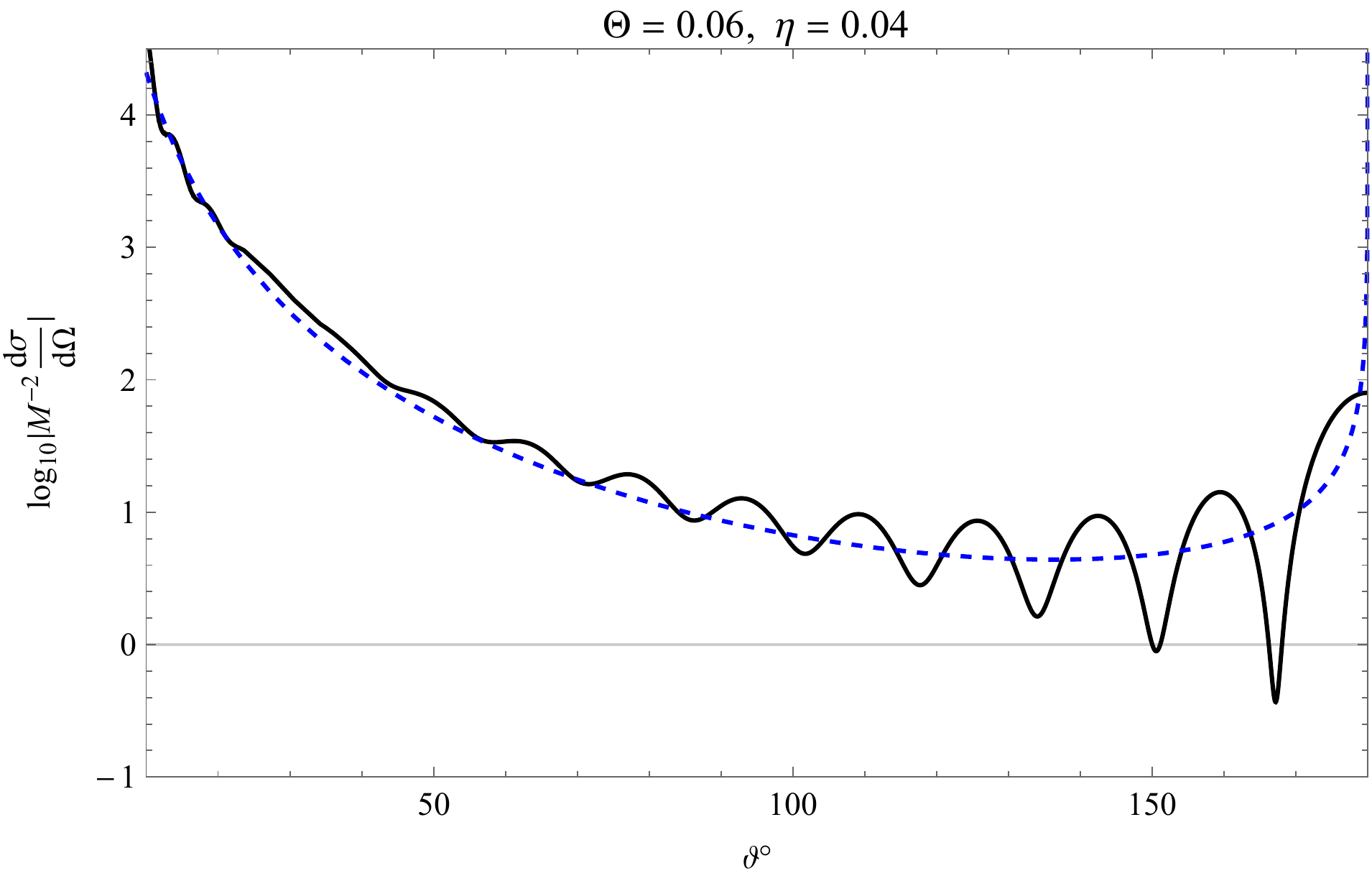}}
  \caption{\footnotesize{Classical and numerical scattering differential cross section. 
  For the numerics we assume $M\omega=1$ in (a) and $M\omega = 2$ in (b).} }
  \label{scatclas2}
\end{figure}

In Fig.~\ref{scatclas2} we show the behavior of the curves for the classical and numerical case. As expected the results for classical analysis fit well in the numerical plot for small angles.

\section{Conclusions}
\label{conc}
In summary, in this paper, we investigate the process of massless scalar wave scattering due to a noncommutative black hole with a global monopole through the partial wave method. 
We have obtained, at the low frequency limit, the phase shift analytically.
The effects of the noncommutativity cause a reduction in the differential scattering/absorption cross
section and shadow radius, while the presence of the global monopole has the effect of increasing the value of such quantities.
In addition, we show that these quantities are non-zero when we take the limit of the zero mass parameter.
Rather,  they are proportional to a minimum mass. 
Moreover, we have also checked the analytically obtained results by numerically solving the radial equation for arbitrary frequencies.

\acknowledgments
We would like to thank CNPq, CAPES and CNPq/PRONEX/FAPESQ-PB (Grant no. 165/2018), for partial financial support. MAA, FAB and EP acknowledge support from CNPq (Grant nos.  306398/2021-4, 312104/2018-9, 304852/2017-1).


\begin{thebibliography}{100}

\bibitem{Hawking:1971vc}
S.~W.~Hawking,
{``\it Black holes in general relativity},''
Commun. Math. Phys. \textbf{25}, 152-166 (1972)
doi:10.1007/BF01877517

\bibitem{Padmanabhan:2003gd}
T.~Padmanabhan,``{\it Gravity and the thermodynamics of horizons},''
Phys. Rept. \textbf{406}, 49-125 (2005)
[arXiv:gr-qc/0311036 [gr-qc]].


\bibitem{Szabo:2006wx}
R.~J.~Szabo,
``{\it Symmetry, gravity and noncommutativity},''
Class. Quant. Grav. \textbf{23}, R199-R242 (2006)
doi:10.1088/0264-9381/23/22/R01
[arXiv:hep-th/0606233 [hep-th]].

\bibitem{Ansoldi:2006vg}
S.~Ansoldi, P.~Nicolini, A.~Smailagic and E.~Spallucci,
{``\it Noncommutative geometry inspired charged black holes},''
Phys. Lett. B \textbf{645}, 261-266 (2007) 
doi:10.1016/j.physletb.2006.12.020
[arXiv:gr-qc/0612035 [gr-qc]].

\bibitem{Nicolini:2008aj}
P.~Nicolini,
``{\it Noncommutative Black Holes, The Final Appeal To Quantum Gravity: A Review},''
Int. J. Mod. Phys. A \textbf{24}, 1229-1308 (2009)
doi:10.1142/S0217751X09043353
[arXiv:0807.1939 [hep-th]].

\bibitem{Nicolini:2005vd}
P.~Nicolini, A.~Smailagic and E.~Spallucci,
``{\it Noncommutative geometry inspired Schwarzschild black hole},''
Phys. Lett. B \textbf{632}, 547-551 (2006)
doi:10.1016/j.physletb.2005.11.004
[arXiv:gr-qc/0510112 [gr-qc]].


\bibitem{Nozari:2008rc}
K.~Nozari and S.~H.~Mehdipour,
{``\it Hawking Radiation as Quantum Tunneling from Noncommutative Schwarzschild Black Hole},''
Class. Quant. Grav. \textbf{25}, 175015 (2008)
doi:10.1088/0264-9381/25/17/175015
[arXiv:0801.4074 [gr-qc]].

\bibitem{Vilenkin:2000jqa}
A.~Vilenkin and E.~P.~S.~Shellard,
{\it Cosmic Strings and Other Topological Defects},
Cambridge University Press, 2000.

\bibitem{Kibble:1976sj}
T.~W.~B.~Kibble,
{ ``\it Topology of Cosmic Domains and Strings},''
J. Phys. A \textbf{9}, 1387-1398 (1976)
doi:10.1088/0305-4470/9/8/029


\bibitem{BezerradeMello:1996si}
E.~R.~Bezerra de Mello and C.~Furtado,
``{\it The Nonrelativistic scattering problem by a global monopole},''
Phys. Rev. D \textbf{56}, 1345-1348 (1997)
doi:10.1103/PhysRevD.56.1345

\bibitem{Pitelli:2009kd}
J.~P.~M.~Pitelli and P.~S.~Letelier,
{\it ``Quantum Singularities Around a Global Monopole},''
Phys. Rev. D \textbf{80}, 104035 (2009)
doi:10.1103/PhysRevD.80.104035
[arXiv:0911.2626 [gr-qc]].

\bibitem{Sharif:2015kna}
M.~Sharif and S.~Iftikhar,
{``\it Null Geodesics and Strong Field Gravitational Lensing of Black Hole with Global Monopole},''
Adv. High Energy Phys. \textbf{2015}, 854264 (2015)
doi:10.1155/2015/854264

\bibitem{Shaikh:2019fpu}
R.~Shaikh,
{\it ``Black hole shadow in a general rotating spacetime obtained through Newman-Janis algorithm},''
Phys. Rev. D \textbf{100}, no.2, 024028  (2019) 
doi:10.1103/PhysRevD.100.024028
[arXiv:1904.08322 [gr-qc]].

\bibitem{Haroon:2019new}
S.~Haroon, K.~Jusufi and M.~Jamil,
{\it ``Shadow Images of a Rotating Dyonic Black Hole with a Global Monopole Surrounded by Perfect Fluid},''
Universe \textbf{6}, no.2, 23 (2020) 
doi:10.3390/universe6020023
[arXiv:1904.00711 [gr-qc]].

\bibitem{Barriola:1989hx}
M.~Barriola and A.~Vilenkin,
{\it ``Gravitational Field of a Global Monopole},''
Phys. Rev. Lett. \textbf{63}, 341 (1989)
doi:10.1103/PhysRevLett.63.341

\bibitem{Anacleto:2017kmg} 
  M.~A.~Anacleto, F.~A.~Brito, S.~J.~S.~Ferreira and E.~Passos,
  {\it ``Absorption and scattering of a black hole with a global monopole in f(R) gravity},''
  Phys.\ Lett.\ B {\bf 788}, 231 (2019)
  doi:10.1016/j.physletb.2018.11.020
  [arXiv:1701.08147 [hep-th]]. 

\bibitem{Townsend:1997ku}
P.~K.~Townsend,
{\it Black holes: Lecture notes},
[arXiv:gr-qc/9707012 [gr-qc]].

\bibitem{Futterman:1988ni}
J.~A.~H.~Futterman, F.~A.~Handler and R.~A.~Matzner,
{\it Scattering from black holes} (Cambridge University Press, England, 1988)
doi:10.1017/CBO9780511735615



\bibitem{Li:2022wzi}
Q.~Li, C.~Ma, Y.~Zhang, Z.~W.~Lin and P.~F.~Duan,
``{\it Shadow, absorption and Hawking radiation of a Schwarzschild black hole surrounded by a cloud of strings in Rastall gravity},''
Eur. Phys. J. C \textbf{82}, no.7, 658 (2022)
doi:10.1140/epjc/s10052-022-10623-3 

\bibitem{Xing:2022emg}
Y.~Xing, Y.~Yang, D.~Liu, Z.~W.~Long and Z.~Xu,
``{\it The ringing of quantum corrected Schwarzschild black hole with GUP},''
Commun. Theor. Phys. \textbf{74}, no.8, 085404 (2022)
doi:10.1088/1572-9494/ac7cdc
[arXiv:2204.11262 [gr-qc]].

\bibitem{Bisnovatyi-Kogan:2022ujt}
G.~S.~Bisnovatyi-Kogan and O.~Y.~Tsupko,
``{\it Analytical study of higher-order ring images of the accretion disk around a black hole},''
Phys. Rev. D \textbf{105}, no.6, 064040 (2022)
doi:10.1103/PhysRevD.105.064040
[arXiv:2201.01716 [gr-qc]].

\bibitem{Zeng:2021dlj}
X.~X.~Zeng, G.~P.~Li and K.~J.~He,
``{ The shadows and observational appearance of a noncommutative black hole surrounded by various profiles of accretions},''
Nucl. Phys. B \textbf{974}, 115639 (2022)
doi:10.1016/j.nuclphysb.2021.115639
[arXiv:2106.14478 [hep-th]].


\bibitem{Mourad:2021qgo}
M.~F.~Mourad and M.~Abdelgaber,
``{\it Spherically symmetric AdS black holes with smeared mass distribution},''
Mod. Phys. Lett. A \textbf{36}, no.05, 2150029 (2021)
doi:10.1142/S0217732321500292

\bibitem{Jha:2021bue}
S.~K.~Jha and A.~Rahaman,
``{\it Lorentz violation and noncommutative effect on superradiance scattering off Kerr-like black hole and on the shadow of it},''
[arXiv:2111.02817 [gr-qc]].

\bibitem{Li:2022jda}
Q.~Li, C.~Ma, Y.~Zhang, Z.~W.~Lin and P.~F.~Duan,
``{\it Gray-body factor and absorption of the Dirac field in ESTGB gravity},''
Chin. J. Phys. \textbf{77}, 1269-1277 (2022)
doi:10.1016/j.cjph.2022.03.027

\bibitem{Gogoi:2022wyv}
D.~J.~Gogoi and U.~D.~Goswami,
``{\it Quasinormal modes and Hawking radiation sparsity of GUP corrected black holes in bumblebee gravity with topological defects},''
JCAP \textbf{06}, no.06, 029 (2022)
doi:10.1088/1475-7516/2022/06/029
[arXiv:2203.07594 [gr-qc]].

\bibitem{Karmakar:2022idu}
R.~Karmakar, D.~J.~Gogoi and U.~D.~Goswami,
``{\it Quasinormal modes and thermodynamic properties of GUP-corrected Schwarzschild black hole surrounded by quintessence},''
[arXiv:2206.09081 [gr-qc]].

\bibitem{Lobos:2022jsz}
N.~J.~L.~S.~Lobos and R.~C.~Pantig,
``{\it Generalized Extended Uncertainty Principle Black Holes: Shadow and lensing in the macro- and microscopic realms},''
[arXiv:2208.00618 [gr-qc]].

\bibitem{Tsupko:2022yzg}
O.~Y.~Tsupko,
``{\it Shape of higher-order photon rings: analytical description with polar curve},''
[arXiv:2208.02084 [gr-qc]].

\bibitem{Zeng:2022fdm}
X.~X.~Zeng, M.~I.~Aslam and R.~Saleem,
``{\it The Optical Appearance of Charged Four-Dimensional Gauss-Bonnet Black Hole with Strings Cloud and Non-Commutative Geometry Surrounded by Various Accretions Profiles},''
[arXiv:2208.06246 [gr-qc]].

\bibitem{Heydari-Fard:2021qdc}
M.~Heydari-Fard and M.~Heydari-Fard,
``{\it Null geodesics and shadow of 4D Einstein\textendash{}Gauss\textendash{}Bonnet black holes surrounded by quintessence},''
Int. J. Mod. Phys. D \textbf{31}, no.08, 2250066 (2022)
doi:10.1142/S0218271822500663
[arXiv:2109.02059 [gr-qc]].


\bibitem{Heydari-Fard:2021pjc}
M.~Heydari-Fard, M.~Heydari-Fard and H.~R.~Sepangi,
``{\it Null geodesics and shadow of hairy black holes in Einstein-Maxwell-dilaton gravity},''
Phys. Rev. D \textbf{105}, no.12, 124009 (2022)
doi:10.1103/PhysRevD.105.124009
[arXiv:2110.02713 [gr-qc]].


\bibitem{Khodadi:2021gbc}
M.~Khodadi, G.~Lambiase and D.~F.~Mota,
``{\it No-hair theorem in the wake of Event Horizon Telescope},''
JCAP \textbf{09}, 028 (2021)
doi:10.1088/1475-7516/2021/09/028
[arXiv:2107.00834 [gr-qc]].

\bibitem{Fathi:2020sfw}
M.~Fathi, M.~Olivares and J.~R.~Villanueva,
``{\it Gravitational Rutherford scattering of electrically charged particles from a charged Weyl black hole},''
Eur. Phys. J. Plus \textbf{136}, no.4, 420 (2021)
doi:10.1140/epjp/s13360-021-01441-9
[arXiv:2009.03404 [gr-qc]].

\bibitem{Chen:2022ngd}
H.~Chen, H.~Hassanabadi, B.~C.~L\"utf\"uo\u{g}lu and Z.~W.~Long,
``{\it Quantum corrections to the quasinormal modes of the Schwarzschild black hole},''
[arXiv:2203.03464 [gr-qc]].



\bibitem{Matzner:1977dn}
R.~A.~Matzner and M.~P.~Ryan,
{\it ``Low Frequency Limit of Gravitational Scattering},''
Phys. Rev. D \textbf{16}, 1636-1642 (1977)
doi:10.1103/PhysRevD.16.1636

\bibitem{Westervelt:1971pm}
P.~J.~Westervelt,
{\it ``Scattering of electromagnetic and gravitational waves by a static gravitational field - comparison between the classical (general-relativistic) and quantum field-theoretic results},''
Phys. Rev. D \textbf{3}, 2319-2324 (1971)
doi:10.1103/PhysRevD.3.2319

\bibitem{Peters:1976jx}
P.~C.~Peters,
{\it ``Differential Cross-Sections for Weak Field Gravitational Scattering},''
Phys. Rev. D \textbf{13}, 775-777 (1976)
doi:10.1103/PhysRevD.13.775

\bibitem{Sanchez:1976fcl}
N.~G.~Sanchez,
{\it ``Scattering of scalar waves from a Schwarzschild black hole},''
J. Math. Phys. \textbf{17}, no.5, 688 (1976) 
doi:10.1063/1.522949  
  
\bibitem{Sanchez:1976xm}
N.~G.~Sanchez,
{\it ``The Wave Scattering Theory and the Absorption Problem for a Black Hole},''
Phys. Rev. D \textbf{16}, 937-945 (1977)
doi:10.1103/PhysRevD.16.937

\bibitem{Sanchez:1977si}
N.~G.~Sanchez,
{\it``Absorption and Emission Spectra of a Schwarzschild Black Hole},''
Phys. Rev. D \textbf{18}, 1030 (1978)
doi:10.1103/PhysRevD.18.1030

\bibitem{Sanchez:1977vz}
N.~G.~Sanchez,
{\it ``Elastic Scattering of Waves by a Black Hole},''
Phys. Rev. D \textbf{18}, 1798 (1978)
doi:10.1103/PhysRevD.18.1798

\bibitem{DeLogi:1977dp}
W.~K.~De Logi and S.~J.~Kovacs,
{\it ``Gravitational Scattering of Zero Rest Mass Plane Waves},''
Phys. Rev. D \textbf{16}, 237-244 (1977)
doi:10.1103/PhysRevD.16.237

\bibitem{Doran:2001ag}
C.~Doran and A.~Lasenby,
{\it ``Perturbation theory calculation of the black hole elastic scattering cross-section},''
Phys. Rev. D \textbf{66}, 024006 (2002)
doi:10.1103/PhysRevD.66.024006
[arXiv:gr-qc/0106039 [gr-qc]].


\bibitem{Dolan:2007ut} 
S.~R.~Dolan, {\it ``Scattering of long-wavelength gravitational waves},'' Phys.\ Rev.\ D {\bf 77}, 044004 (2008) 
[arXiv:0710.4252 [gr-qc]].

\bibitem{Crispino:2009ki} 
  L.~C.~B.~Crispino, S.~R.~Dolan and E.~S.~Oliveira,
  {\it ``Scattering of massless scalar waves by Reissner-Nordstrom black holes},''
  Phys.\ Rev.\ D {\bf 79}, 064022 (2009)
  doi:10.1103/PhysRevD.79.064022
  [arXiv:0904.0999 [gr-qc]].


\bibitem{Starobinskil:1974nkd}
A.~A.~Starobinskil and S.~M.~Churilov,
{\it ``Amplification of electromagnetic and gravitational waves scattered by a rotating ''black hole''},''
Sov. Phys. JETP \textbf{65}, no.1, 1-5 (1974).

\bibitem{Gibbons:1975kk}
G.~W.~Gibbons,
{\it ``Vacuum Polarization and the Spontaneous Loss of Charge by Black Holes},''
Commun. Math. Phys. \textbf{44}, 245-264 (1975)
doi:10.1007/BF01609829

\bibitem{Page:1976df}
D.~N.~Page,
{\it``Particle Emission Rates from a Black Hole: Massless Particles from an Uncharged, Nonrotating Hole},''
Phys. Rev. D \textbf{13}, 198-206 (1976)
doi:10.1103/PhysRevD.13.198

\bibitem{Churilov1973}  A. A. Starobinskii and S. M. Churilov, Zh. Eksp. Teor. Fiz. {\bf 65}, 3 (1973).

\bibitem{Moura:2011rr} 
  F.~Moura,
  {\it ``Scattering of spherically symmetric $d$-dimensional $\alpha'-$corrected black holes in string theory},''
  JHEP {\bf 1309}, 038 (2013)
  doi:10.1007/JHEP09(2013)038
  [arXiv:1105.5074 [hep-th]].


\bibitem{Jung:2004yh}
E.~Jung and D.~K.~Park,
{\it ``Effect of scalar mass in the absorption and emission spectra of Schwarzschild black hole},''
Class. Quant. Grav. \textbf{21}, 3717-3732 (2004)
doi:10.1088/0264-9381/21/15/007
[arXiv:hep-th/0403251 [hep-th]].

\bibitem{Jung:2004yn}
E.~Jung, S.~Kim and D.~K.~Park,
{\it ``Proof of universality for the absorption of massive scalar by the higher-dimensional Reissner-Nordstrom black holes},''
Phys. Lett. B \textbf{602}, 105-111 (2004)
doi:10.1016/j.physletb.2004.09.067
[arXiv:hep-th/0409145 [hep-th]].


\bibitem{Doran:2005vm}
C.~Doran, A.~Lasenby, S.~Dolan and I.~Hinder,
{\it ``Fermion absorption cross section of a Schwarzschild black hole},''
Phys. Rev. D \textbf{71}, 124020 (2005)
doi:10.1103/PhysRevD.71.124020
[arXiv:gr-qc/0503019 [gr-qc]].

\bibitem{Dolan:2006vj}
S.~Dolan, C.~Doran and A.~Lasenby,
{\it ``Fermion scattering by a Schwarzschild black hole},''
Phys. Rev. D \textbf{74}, 064005 (2006)
doi:10.1103/PhysRevD.74.064005
[arXiv:gr-qc/0605031 [gr-qc]].

\bibitem{Castineiras:2007ma}
J.~Castineiras, L.~C.~B.~Crispino and D.~P.~M.~Filho,
{\it ``Source coupled to the massive scalar field orbiting a stellar object},''
Phys. Rev. D \textbf{75}, 024012 (2007)
doi:10.1103/PhysRevD.75.024012


\bibitem{Benone:2014qaa} 
  C.~L.~Benone, E.~S.~de Oliveira, S.~R.~Dolan and L.~C.~B.~Crispino,
  {\it ``Absorption of a massive scalar field by a charged black hole},''
  Phys.\ Rev.\ D {\bf 89}, no. 10, 104053 (2014)
  doi:10.1103/PhysRevD.89.104053
  [arXiv:1404.0687 [gr-qc]].
   
\bibitem{Marinho:2016ixt} 
  C.~I.~S.~Marinho and E.~S.~de Oliveira,
  ``{\it Scattering of massless scalar waves from Schwarzschild-Tangherlini black holes on the brane},''
  arXiv:1612.05604 [gr-qc].


\bibitem{Das:1996we} 
  S.~R.~Das, G.~W.~Gibbons and S.~D.~Mathur,
  {\it ``Universality of low-energy absorption cross-sections for black holes},''
  Phys.\ Rev.\ Lett.\  {\bf 78}, 417 (1997)
  doi:10.1103/PhysRevLett.78.417
  [hep-th/9609052]. 
     

\bibitem{Macedo:2016yyo} 
  C.~F.~B.~Macedo, L.~C.~B.~Crispino and E.~S.~de Oliveira,
  {\it ``Scalar waves in regular Bardeen black holes: Scattering, absorption and quasinormal modes},''
  Int.\ J.\ Mod.\ Phys.\ D {\bf 25}, no. 09, 1641008 (2016)
  doi:10.1142/S021827181641008X
  [arXiv:1605.00123 [gr-qc]].

\bibitem{deOliveira:2018kcq} 
  E.~S.~de Oliveira,
  {\it ``Scalar scattering from black holes with tidal charge},''
  Eur.\ Phys.\ J.\ C {\bf 78}, no. 11, 876 (2018)
  doi:10.1140/epjc/s10052-018-6316-9
  [arXiv:1805.04987 [gr-qc]].
  
\bibitem{Hai:2013ara} 
  H.~Hai, W.~Yong-Jiu and C.~Ju-Hua,
  {\it ``Absorption cross section of black holes with global monopole},''
  Chin.\ Phys.\ B {\bf 22}, no. 7, 070401 (2013). 



\bibitem{Crispino:2007zz} 
  L.~C.~B.~Crispino, E.~S.~Oliveira and G.~E.~A.~Matsas,
  {\it ``Absorption cross section of canonical acoustic holes},''
  Phys.\ Rev.\ D {\bf 76}, 107502 (2007).

\bibitem{Dolan:2009zza} 
  S.~R.~Dolan, E.~S.~Oliveira and L.~C.~B.~Crispino,
  {\it ``Scattering of sound waves by a canonical acoustic hole},''
  Phys.\ Rev.\ D {\bf 79}, 064014 (2009)

\bibitem{Oliveira:2010zzb} 
  E.~S.~Oliveira, S.~R.~Dolan and L.~C.~B.~Crispino,
  {\it ``Absorption of planar waves in a draining bathtub},''
  Phys.\ Rev.\ D {\bf 81}, 124013 (2010).
 

\bibitem{Dolan:2011zza} 
  S.~R.~Dolan, E.~S.~Oliveira and L.~C.~B.~Crispino,
  {\it ``Aharonov-Bohm effect in a draining bathtub vortex},''
  Phys.\ Lett.\ B {\bf 701}, 485 (2011).
  doi:10.1016/j.physletb.2011.06.013
  
\bibitem{Anacleto:2012ba}
M.~A.~Anacleto, F.~A.~Brito and E.~Passos,
{\it``Analogue Aharonov-Bohm effect in a Lorentz-violating background},''
Phys. Rev. D \textbf{86}, 125015 (2012)
doi:10.1103/PhysRevD.86.125015
[arXiv:1208.2615 [hep-th]].  
  
\bibitem{Anacleto:2012du}
M.~A.~Anacleto, F.~A.~Brito and E.~Passos,
{\it``Noncommutative analogue Aharonov-Bohm effect and superresonance},''
Phys. Rev. D \textbf{87}, no.12, 125015 (2013)
doi:10.1103/PhysRevD.87.125015
[arXiv:1210.7739 [hep-th]]. 



\bibitem{Dolan:2012yc} 
  S.~R.~Dolan and E.~S.~Oliveira,
  {\it ``Scattering by a draining bathtub vortex},''
  Phys.\ Rev.\ D {\bf 87}, no. 12, 124038 (2013)
  doi:10.1103/PhysRevD.87.124038
  [arXiv:1211.3751 [gr-qc]].  
  
\bibitem{Anacleto:2015mta} 
  M.~A.~Anacleto, I.~G.~Salako, F.~A.~Brito and E.~Passos,
  {\it ``Analogue Aharonov-Bohm effect in neo-Newtonian theory},''
  Phys.\ Rev.\ D {\bf 92}, no. 12, 125010 (2015)
  doi:10.1103/PhysRevD.92.125010
  [arXiv:1506.03440 [hep-th]]; 

\bibitem{Anacleto:2016ukc}
M.~A.~Anacleto, F.~A.~Brito, A.~Mohammadi and E.~Passos,
{\it ``Aharonov-Bohm effect for a fermion field in the acoustic black hole ''spacetime''},''
Eur. Phys. J. C \textbf{77}, no.4, 239 (2017)
doi:10.1140/epjc/s10052-017-4801-1
[arXiv:1606.09231 [hep-th]].

 
\bibitem{Anacleto:2014cga}
M.~A.~Anacleto, F.~A.~Brito and E.~Passos,
{\it ``Gravitational Aharonov\textendash{}Bohm effect due to noncommutative BTZ black hole},''
Phys. Lett. B \textbf{743}, 184-188 (2015)
doi:10.1016/j.physletb.2015.02.056
[arXiv:1408.4481 [hep-th]]
  
\bibitem{Anacleto:2019tdj} 
  M.~A.~Anacleto, F.~A.~Brito, J.~A.~V.~Campos and E.~Passos,
  {\it ``Absorption and scattering of a noncommutative black hole},''
Phys.\ Lett.\ B {\bf 803}, 135334 (2020)
  doi:10.1016/j.physletb.2020.135334  
 [arXiv:1907.13107 [hep-th]].

\bibitem{Anacleto:2018acl}
M.~A.~Anacleto, F.~A.~Brito, J.~A.~V.~Campos and E.~Passos,
{\it ``Higher-derivative analogue Aharonov\textendash{}Bohm effect, absorption and superresonance},''
Int. J. Mod. Phys. A \textbf{35}, no.21, 2050112 (2020)
doi:10.1142/S0217751X20501122
[arXiv:1810.13356 [hep-th]].

\bibitem{Anacleto:2020lel}
M.~A.~Anacleto, F.~A.~Brito, J.~A.~V.~Campos and E.~Passos,
{\it ``Quantum-corrected scattering and absorption of a Schwarzschild black hole with GUP},''
Phys. Lett. B \textbf{810}, 135830 (2020)
doi:10.1016/j.physletb.2020.135830
[arXiv:2003.13464 [gr-qc]].

\bibitem{Anacleto:2020zhp}
M.~A.~Anacleto, F.~A.~Brito, J.~A.~V.~Campos and E.~Passos,
{\it ``Absorption and scattering by a self-dual black hole},''
Gen. Rel. Grav. \textbf{52}, no.10, 100 (2020)
doi:10.1007/s10714-020-02756-1
[arXiv:2002.12090 [hep-th]].


\bibitem{Anacleto:2020efy}
M.~A.~Anacleto, F.~A.~Brito, B.~R.~Carvalho and E.~Passos,
{\it ``Noncommutative correction to the entropy of BTZ black hole with GUP},''
Adv. High Energy Phys. \textbf{2021}, 6633684 (2021)
doi:10.1155/2021/6633684
[arXiv:2010.09703 [hep-th]].

\bibitem{Anacleto:2020zfh}
M.~A.~Anacleto, F.~A.~Brito, S.~S.~Cruz and E.~Passos,
{\it ``Noncommutative correction to the entropy of Schwarzschild black hole with GUP},''
Int. J. Mod. Phys. A \textbf{36}, no.03, 2150028 (2021)
doi:10.1142/S0217751X21500287
[arXiv:2010.10366 [hep-th]].


\bibitem{Campos:2021sff}
J.~A.~V.~Campos, M.~A.~Anacleto, F.~A.~Brito and E.~Passos,
{\it ``Quasinormal modes and shadow of noncommutative black hole},''
Sci. Rep. \textbf{12}, no.1, 8516 (2022)
doi:10.1038/s41598-022-12343-w
[arXiv:2103.10659 [hep-th]].

  
\bibitem {Yennie1954} 
D.~R.~Yennie, D.~G.~Ravenhall and R.~N.~Wilson,
``{\it Phase-Shift Calculation of High-Energy Electron Scattering},''
Phys. Rev. \textbf{95}, 500-512 (1954)
doi:10.1103/PhysRev.95.500 
 

\bibitem{Cotaescu:2014jca} 
  I.~I.~Cotaescu, C.~Crucean and C.~A.~Sporea,
  {\it ``Partial wave analysis of the Dirac fermions scattered from Schwarzschild black holes},''
  Eur.\ Phys.\ J.\ C {\bf 76}, no. 3, 102 (2016)
  doi:10.1140/epjc/s10052-016-3936-9
  [arXiv:1409.7201 [gr-qc]]. 
  




\end{thebibliography}
\end{document}